\documentclass[10pt, conference, letterpaper]{IEEEtran} 
\IEEEoverridecommandlockouts

\usepackage{ifthen}

\usepackage[tbtags]{amsmath} 
\usepackage{amssymb}  
\usepackage{verbatim} 
\usepackage{amsxtra}  

\usepackage{mathabx}

\usepackage{enumerate}

\usepackage{amsfonts}
\usepackage{color}

\usepackage{caption}
\usepackage{subcaption}

\usepackage{wasysym}

\usepackage{cite}

\usepackage{mathtools}
\mathtoolsset{showonlyrefs=true}

\usepackage{tikz}

\usetikzlibrary{shapes,arrows,decorations.markings,positioning}

\tikzstyle{plant} = [draw, fill=red!5, rounded rectangle, 
    minimum height=3em]
\tikzstyle{block} = [draw, fill=blue!5, rectangle, 
    minimum height=3em]
\tikzstyle{tap} = [draw, fill=olive!10, rectangle, minimum height=3em]
\tikzstyle{sum} = [draw, fill=yellow!10, circle, node distance=1cm]
\tikzstyle{pinstyle} = [pin edge={to-,thick,black}]

\tikzstyle{BitPipe} = [thick, decoration={markings,mark=at position
   1 with {\arrow[semithick]{triangle 60}}},
   double distance=1.4pt, shorten >= 5.5pt,
   preaction = {decorate},
   postaction = {draw,line width=1.4pt, black, shorten >= 4.5pt}]

\usetikzlibrary{chains,shapes.multipart}
\tikzstyle{FIFO} = [rectangle split, rectangle split parts=3, draw, rectangle split horizontal,minimum height=3em,text height=1.5em,text depth=1em,on chain,inner ysep=0pt]

\usetikzlibrary{fit,backgrounds}

\tikzstyle{coord} = [coordinate]
\tikzstyle{gain} = [draw, fill=red!5, regular polygon, regular polygon sides=3, shape border rotate=-90]

\tikzset{mwe/.style={minimum size=0.5em}}

\usepackage{epsfig, graphicx, psfrag}

\usepackage[mathcal]{euscript}
\usepackage[cal=boondox]{mathalfa}

\usepackage{amsthm}

\newtheorem{thm}{Theorem}[section]

\newtheorem{cor}{Corollary}[section]

\theoremstyle{definition}

\newtheorem*{defn*}{Definition}

\newtheorem*{scheme*}{Scheme}

\theoremstyle{remark}
\newtheorem{remark}{Remark}[section]

\providecommand{\thmref}[1]{Th.~\ref{#1}}

\providecommand{\secref}[1]{Sec.~\ref{#1}}

\providecommand{\propref}[1]{Prop.~\ref{#1}}
\providecommand{\remref}[1]{Rem.~\ref{#1}}
\providecommand{\figref}[1]{Fig.~\ref{#1}}
\providecommand{\corref}[1]{Corol.~\ref{#1}}

\newcommand{\ie}{i.e.}
\newcommand{\iid}{i.i.d.}
\newcommand{\eg}{e.g.}

\newcommand{\cf}{cf.}

\newcommand{\ints}{\mathbb{Z}}

\newcommand{\nats}{\mathbb{N}}
\newcommand{\rats}{\mathbb{Q}}

\newcommand{\mI}{\mathcal{I}}

\newcommand{\cQ}{\mathcal{Q}}
\newcommand{\Comment}[1]{}
\newcommand{\old}[1]{}
\newcommand{\rem}[1]{}

\newcommand{\veps}{\varepsilon}

\newcommand{\hW}{\hat{W}}

\newcommand{\tQ}{\tilde{Q}}

\newcommand{\bp}{\mathbf{p}}
\newcommand{\bP}{\mathbf{P}}
\newcommand{\bQ}{\mathbf{Q}}
\newcommand{\tbQ}{\tilde{\bQ}}

\providecommand{\comment}[1]{}

\newcommand{\beqn}[1]{\begin{eqnarray}\label{#1}}
\newcommand{\eeqn}{\end{eqnarray}}
\newcommand{\beq}[1]{\begin{equation}\label{#1}}
\newcommand{\eeq}{\end{equation}}

\makeatletter
\newcommand{\bigger}{\bBigg@{3.2}}
\newcommand{\vast}{\bBigg@{4}}
\newcommand{\Vast}{\bBigg@{5}}
\makeatother

\providecommand{\KL}[2]{{\bbD \left( #1 \middle\| #2 \right)}}

\providecommand{\bbD}{\mathbb{D}}
\providecommand{\bbE}{\mathbb{E}}

\providecommand{\E}[1]{\bbE \left[ #1 \right]}

\newcommand{\PR}[1]{\Pr\left( #1 \right)}
\newcommand{\CPR}[2]{\Pr\left( #1 \middle| #2 \right)}

\providecommand{\BLER}[1]{P_e \left( #1 \right)}
\providecommand{\Eblk}{E}

\providecommand{\Exp}[1]{\exp \left\{ #1 \right\}}
\providecommand{\IV}{\vec{V}}
\providecommand{\Hb}[1]{H_b \left( #1 \right)}

\usepackage{mathtools}

\mathtoolsset{showonlyrefs=true}

\usepackage{MnSymbol}

\usepackage{comment}

\newcommand{\VersionLength}{short}

\providecommand{\ver}{\ifthenelse{\equal{\VersionLength}{long}}}
\providecommand{\nver}{\ifthenelse{\equal{\VersionLength}{short}}}

\providecommand{\figref}[1]{Fig.~\ref{#1}}
\providecommand{\secref}[1]{Sec.~\ref{#1}}

\ver{}{
	\setlength{\abovedisplayskip}{3.5pt}
	\setlength{\belowdisplayskip}{4pt}

}

\ver{\usepackage{subfig}}{}
\ver{\usepackage{hyperref}}{\usepackage{url}}

\begin{document}

\title{The Information Velocity of Packet-Erasure Links}

\author{Elad Domanovitz, Tal Philosof, and Anatoly Khina
    \thanks{This work was supported by 
    the \textsc{Israel Science Foundation} (grant No.\ 2077/20) 
    and by the WIN Consortium through the Israel Innovation Authority.}
    \thanks{E.~Domanovitz and A.~Khina are with the Department of Electrical Engineering--Systems, Tel Aviv University, Tel Aviv~6997801, Israel (e-mails: \texttt{domanovi@mail.tau.ac.il}, \texttt{anatolyk@eng.tau.ac.il}).}
    \thanks{T.~Philosof is with Samsung Research, Tel Aviv 6492103, Israel (e-mail: \texttt{tal.philosof@gmail.com})}
}
\maketitle




\begin{abstract}
    We consider the problem of in-order packet transmission
    over a cascade of packet-erasure links with acknowledgment (ACK) signals, interconnected by relays.
    We treat first the case of transmitting a single packet, in which ACKs are unnecessary,
    over links with independent identically distributed erasures.
    For this case, we derive tight upper and lower bounds on the probability of arrive failure within an allowed end-to-end communication delay over a given number of links.
    When the number of links is commensurate with the allowed delay, we determine the maximal ratio between the two---coined information velocity---for which the arrive-failure probability decays to zero;
    we further derive bounds on the arrive-failure probability when the ratio is below the information velocity, determine the exponential arrive-failure decay rate, and extend the treatment to links with different erasure probabilities. 
    We then elevate all these results for a stream of packets with independent geometrically distributed interarrival times, and prove that the information velocity and the exponential decay rate remain the same for any stationary ergodic arrival process and for deterministic interarrival times.
    We demonstrate the significance of the derived fundamental limits---the information velocity and the arrive-failure exponential decay rate---by comparing them to simulation results.
\end{abstract}


\allowdisplaybreaks


\section{Introduction}
\label{s:intro}

Wireless communications technologies are gradually shifting toward working over distributed network topologies 
of increasing numbers of smaller-size units, to allow lower energy consumption, reduced delay, and ubiquitous connectivity. 

Examples of such communications technologies include the following. Cellular vehicle-to-everything (C-V2X) is gaining prominence in fifth generation (5G) cellular technology and will play a major role in future generations \cite{lien20203gpp}. For applications such as emergency vehicle coordination and platooning, C-V2X requires highly-reliable packet transmission with low latency. In addition, C-V2X, including the enhancement of user-equipment (UE) relaying, needs to support hundreds of nodes comprising vehicles, pedestrian and road units, in highly congested areas. The end-to-end (E2E) latency and velocity with which information propagates to distant vehicles are key performance indicators of such system. Device-to-device (D2D) plays a similar role in emergency management systems.

Another key technology introduced in 5G is the Internet of Things (IoT). 
IoT allows to connect billions of physical devices around the world, all collecting, monitoring, processing, and sharing data by relying on the ubiquity of wireless connectivity. However, since replenishing of many of the IoT units is challenging, 
such units have severe energy limitations posed by their limited battery capacity.
To alleviate this problem, machine-to-machine (M2M) communications is being promoted as a tool to forward information across large distances from end units to the core network via other agents, which act as relays.

Without delay constraints, the maximal reliable communication rate---Shannon's capacity \cite[Ch.~7]{CoverBook2Edition}---over a cascade of a finite number of links interconnected by relays is equal to the minimum of the individual link capacities. 

However, less is clear about the error-probability exponent (EE) at rates below the capacity in this setting, 
when the nodes apply causal operations to their measurements.
In fact, even determining the EE of transmitting a \textit{single bit} over a tandem of binary symmetric channels is a challenging problem \cite{Huleihel-Polyanskiy-Shayevitz:real-time-relaying:ISIT2019,Jog-Loh:Real-Time-Relaying:IT2020}, 
which was only recently resolved by Ling and Scarlett~\cite{Ling-Scarlett:Real-Time-Relaying:ISIT2021,Ling-Scarlett:Real-Time-Relaying:Arxiv2021}, 
who showed that 
the EE of transmitting a single bit is equal to the minimum of the corresponding individual EEs. 

Moreover, when the number of links is commensurate with the number of transmit time steps,
even the capacity is not known and depends on the ratio between the number of links and transmit time steps, when the two grow to infinity. 
In fact, even the maximal possible ratio in transmitting a single bit with arbitrarily small error probability---termed \textit{Information Velocity} by Polyanskiy (see \cite{Huleihel-Polyanskiy-Shayevitz:real-time-relaying:ISIT2019}; see also \cite{Rajagopalan-Schulman:FOCS:Real-Time-Networks:STOC1994})---is yet to be determined; the same term was used earlier by Iyer and Vaze \cite{Information-Velocity:Wireless:WiOpt2015} in a related setting of spatial wireless networks, as well as in other disciplines, \eg, in physics, in neuroscience, epidemic spread in networks,
and in marketing and finance. For packet-erasure links, much work has been done on analyzing the E2E delay of transmitted packets over multi-link networks for various setting; see, \eg,  \cite{TandemQueues:Qos:MobileComputing2008,Xie-Haenggi:Multihop:E2E-Delay:AdHoc2009,Gupta-Shroff:Multihop:delay:INFOCOM2009,Dikaliotis-Dimakis-Alexandros-Ho-Effros:Multihop:delay:IT2014,PacketLinkMeasure:VTC2008,Vellambi-Torabkhani-Fekri:MultiHop:Delay:IT2011,Kayl-Yates-Gruteser:WhenToUpdate:INFOCOM2012,MultihopPAN:VTC2014,Yang-Haenggi-Xiao:MultiHop:delay:COM2018,Yates:LIFO:INFOCOM2018,Devassy-Durisi-Ferrante-Simeone-Uysal:Queue:JSAC2019,Soret-Ravikanti-Popovski:MultiHop:delay:ICC2020,jacquet2010information} and the references therein. 

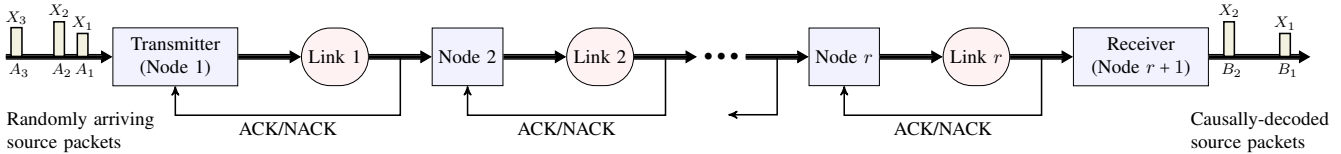
\begin{figure*}[!th]
    \resizebox{\textwidth}{!}{\begin{tikzpicture}[auto, arrow/.style={very thick, ->, >=stealth'},start chain=going right,>=latex,node distance=.13\columnwidth,>=latex']
    \node[coordinate] (input) {};
    \node[block, right of = input, node distance = .35\columnwidth] (enc) {\begin{tabular}{c} Transmitter \\ (Node 1) \end{tabular}};
    \node[plant, right = of enc] (link1) {Link 1};
    \node[block, right = of link1] (node2) {Node 2};
    \node[plant, right = of node2] (link2) {Link 2};
    \node[mwe, right = of link2] (etc) {$\bullet\bullet\bullet$};
    \node[block, right = of etc] (nodeR) {Node $r$};
    \node[plant, right = of nodeR] (linkR) {Link $r$};
    \node[block, right = of linkR] (dec) {\begin{tabular}{c} Receiver \\ (Node $r+1$) \end{tabular}};
    \node[coordinate, right of = dec, node distance = .35\columnwidth] (output) {};
    \node[mwe, below of = etc, node distance = 10.6mm] (FBr) {};

    \draw[BitPipe] (input) -- node [tap, pos=0.1, scale=0.5, minimum width= .4cm, above] {}  node [pos=0.1, below] {\footnotesize{ $A_3$}} node[pos=.1, above, minimum height = 14mm] {\footnotesize $X_3$}  
    node [tap, pos=0.5, scale = 0.6, minimum width = 0.32cm, above] {} node [pos=0.5, below] {\footnotesize{ $A_2$}} node[pos=.5, above, minimum height = 16.5mm] {\footnotesize $X_2$}
    node [tap, pos=0.72, scale=0.4, minimum width = .485cm, above] {} node [pos=0.72, below] {\footnotesize{ $A_1$}} node[pos=.72, above, minimum height = 12mm] {\footnotesize $X_1$}
    node[below, pos=.7]{\begin{tabular}{l}
          \\ \\
          Randomly arriving
       \\ source packets
     \end{tabular}}
    (enc);

    \draw[BitPipe] (dec) -- 
    node [tap, pos=0.2, scale=0.6, minimum width= .32cm, above] {}  node [pos=0.2, below] {\footnotesize{ $B_2$}}
    node[pos=.2, above, minimum height = 16.5mm]{\footnotesize $X_2$} 
    node [tap, pos=0.75, scale=0.4, minimum width = .485cm, above] {} node [pos=0.75, below] {\footnotesize{ $B_1$}} node[pos=.75, above, minimum height = 12mm] {\footnotesize $X_1$}
    node [below, pos=0.5] {\begin{tabular}{l} 
                             \\ \\ 
                                Causally-decoded 
                             \\ source packets 
                           \end{tabular}} 
                           (output);

    \draw[BitPipe] (enc) -- (link1);
    \draw[BitPipe] (link1) -- node (y1) {} (node2);
    \draw[BitPipe] (node2) -- (link2);
    \draw[BitPipe] (link2) -- node (y2) {} (etc);
    \draw[BitPipe] (etc) -- node (yR) {} (nodeR);
    \draw[BitPipe] (nodeR) -- (linkR);
    \draw[BitPipe] (linkR) -- node (yRx) {} (dec);

    \draw[arrow,thick] (y1) to++(0,-12mm)-| node [below,pos=.25] {ACK/NACK} (enc);
    \draw[arrow,thick] (y2) to++(0,-12mm)-| node [below,pos=.25] {ACK/NACK} (node2);
    \draw[arrow,thick] (yR) |- (FBr);
    \draw[arrow,thick] (yRx) to++(0,-12mm)-| node [below,pos=.25] {ACK/NACK} (nodeR);
\end{tikzpicture}}
    \caption{Block diagram of the system model. $X_i$ denotes the content of packet $i$.} 
    \label{fig:relays}
\end{figure*}

For packet-based networks, the scaling of the E2E delay as a function of the number of servers in cascade with various service distributions (where the communications between links is assumed to have zero delay and zero errors) was analyzed in previous works: for deterministic arrival and service curve models, deterministic min--plus algebra has been used to show that worst-case E2E delays grow linearly in the number of nodes \cite{le2001network}. Later, linear growth of E2E delay was shown also for the case of stochastic arrival and service curves (see, e.g., \cite{Fidler2010SurveyOfDetAndStochService} for a survey of deterministic and stochastic service curve models) using stochastic network calculus (see, e.g., \cite{Fidler2015aGuideToTheStichNetCalc}), while assuming that the service at different nodes is statistically independent \cite{chang2000performance, Fidler2006endtoendProb}.

In this work, we concentrate on in-order communication over a cascade of packet-erasure links with acknowledgment (ACK) feedback, depicted in \figref{fig:relays}. 
The erasures in each link are assumed independent and identically distributed (\iid). 

For this setting, we derive upper and lower bounds on the arrive-failure probability of a packet over $r$ relays during $N$ time steps, namely, the probability of the E2E delay of a packet to exceed an allowed delay threshold $N$.
Beyond being important on their own right, these bounds allow us, in turn, to determine the fundamental limits of this problem where the ratio between $r$ and $N$ is held constant: 
\begin{itemize}
\item 
    The \textit{information velocity (IV)}---the maximal speed that information can spread reliably across a cascade of links. 
\item 
    The \textit{error exponent (EE)}---The exponential decay of the arrive-failure probability when $r/N$ is below the IV.
\end{itemize}
While the scaling of the E2E delay was known to be linear for independent service time, 
the IV constitutes the best exact asymptotic scaling rate at which reliable communication may be attained.
Alternatively, the IV supplements and extends the single-link capacity to multi-hop scenarios (see \secref{ss:discussion:1packet-r-links_vs_r-packets-1link}).
The EE further quantifies the exact exponential decay rate when transmitting below the IV threshold.

To derive closed-form expressions for these quantities, we focus on a simple (yet relevant) model of arrival and service. While these results might be derived from existing bounds on the arrive-failure probability for more general arrival and service models, 
to the best of our knowledge, it is the first time the IV and EE are explicitly described for a non-trivial setting.

Interestingly, when considering pure queueing-theory scenarios, the independence assumption of service times is hard to justify \cite{burchard2010superlinear}. The problem formulation we suggest provides a compelling justification for the independence assumption from a communication prospective. 

The rest of the paper is organized as follows.
Secs.~\ref{ss:intro:notation} and \ref{s:model} describe the used notation and the communication setup that is treated in this work, respectively.
\secref{s:1packet} considers the case of single-packet transmission: 
upper and lower bounds on the arrive-failure probability are derived for homogeneous links (all links having the same erasure probability), 
and the IV and EE are determined for this setting in \secref{ss:IV:same-channel}; 
this treatment is then extended to heterogeneous links in \secref{ss:IV:different-channels}. 
The single-packet results are then elevated to a stream of causally arriving packets to the transmitter by tools from queueing theory in \secref{s:lambda>0}: first, for an \iid\ Bernoulli arrival process (equivalently, a process with \iid\ geometric interarrival times) in \secref{ss:lambda>0:Bernoulli}; the IV and the EE are then shown to remain the same for any ergodic arrival process as well as for deterministic interarrival times with the same arrival rate, in \secref{ss:lambda>0:non-Bernoulli}.
Numerical results are presented in \secref{s:numeric}, which confirm the usefulness of the derived theoretical results.
\secref{s:extensions} offers alternative plausible definitions of the IV, along with extensions to single-packet communication without feedback (akin to the definition in \cite{Huleihel-Polyanskiy-Shayevitz:real-time-relaying:ISIT2019}) and instantaneous~links. The paper is concluded by a discussion of the IV and Shannon's capacity, the in-order transmission assumption, and anytime (and anywhere) reliability of networked control \cite{SahaiMitterPartI,Sahai:NotTheSame:IT2008}.


\subsection{Notation}
\label{ss:intro:notation}

$\nats, \ints$, and $\rats$ denote the sets of natural, integeres, and rational numbers, respectively, and $[n] \triangleq \{1, 2, \ldots, n\}$ denotes the smallest $n \in \nats$ natural numbers. $\lceil \cdot \rceil$ and $\lfloor \cdot \rfloor$ denote the floor and ceiling operations, respectively. Vectors are denoted by boldfaced letters ($\mathbf{x}$).
The standard $n$-simplex for $n \in \nats$ is denoted by $\Delta_n \triangleq \left\{ (x_1, x_2, \ldots, x_{n+1}) \middle| \sum_{\ell=1}^{n+1} x_\ell = 1; x_\ell \geq 0, \ell \in [n+1] \right\}$.
$\log$ and $\exp$ denote the logarithm and exponentiation operations and are understood to the same base. We denote by $\Hb{p} \triangleq -p \log p - (1-p) \log (1-p)$ and $\KL{p}{q} \triangleq p \log \frac{p}{q} + (1-p) \log \frac{1-p}{1-q}$
the binary entropy and the binary Kullback--Leibler (KL) divergence, respectively, for $p, q \in [0,1]$, with the convention that $0 \log 0 \triangleq 0$, $0 \log \frac{0}{0} \triangleq 0$, and $ \log \frac{1}{0} \triangleq \infty$. We use standard $o$-notation; in particular, $g(N) = o(N)$ and $f(N) = o(1)$ mean that $\lim\limits_{N \to \infty} g(N)/N = \lim\limits_{N \to \infty} f(N) = 0$.


\section{Communication Setup}
\label{s:model}

The communication model that we consider in this work is detailed next and is depicted in \figref{fig:relays}.

\textit{Source Stream.}
We consider the transmission of a stream of source packets, 
with packet $i \in \ints$ arriving at time $A_i \in \ints$. We further define the $i$-th interarrival time by $D_i = A_i - A_{i-1}$.

\textit{Cascade of erasure links.}
We will consider cascades of $r$ independent links: The output of link $i \in [r]$ servers as the input to node $i+1$. 
Node $i$ transmits its received packets \textit{in order} over link $i$.
At each time step, a transmitted packet by node $i$ over link $i$ arrives to its destination with probability $1-p_i$
and is erased with probability $p_i \in [0, 1)$. Arrived packets are \textit{acknowledged} meaning that node $i$ knows when its transmitted packet arrived successfully to node $i+1$ in the previous time step;
in case of an erasure, node $i$ retransmits the same packet over subsequent time steps, until it successfully arrives to the next node (and acknowledged).
The erasure events over link $i$ are assumed \iid;
the links' independence means that the erasure events across the different links are mutually independent as well. 

\textit{Departure Process.} 
Denote by $B_i$ the time at which source packet $i$ arrives to the end receiver (node $r+1$). 
Clearly $B_i \geq A_i + r$ since each link causes a loss of at least one time unit.\footnote{See \secref{ss:extensions:1packet:instant-hops} for the setting of ``instantaneous links''.}

The following notions will be considered in this work.

\textit{Arrive-Failure Probability.}
Denote by $N$ the allowed E2E delay for each packet.
Then, the arrive-failure probability is defined as the probability of the E2E delay to exceed this allowed value:
\begin{align}
\label{eq:failure-prob}
    \BLER{N} \triangleq \sup_{i \in \mI \subseteq \ints} \PR{B_i > A_i + N}, 
\end{align}
where
$\mI = \{1\}$ for single-packet transmission (\secref{s:1packet}), $\mI = \ints$ for steady-state multi-packet transmission (\secref{s:lambda>0}), and $\mI = \{1, 2, \ldots m\}, m \in \nats$ for (transient) $m$-packet transmission.

\textit{Information Velocity.}
Let $r$ grow linearly with $N$ at $\alpha$ ratio:
\begin{align} 
\label{eq:r=alphaN}
    r = \left\lceil \alpha N \right\rceil, 
\end{align} 
such that the arrive-failure probability $\BLER{N}$ of \eqref{eq:failure-prob}
decays to zero with $N$. 
The largest possible such $\alpha$ will be referred to as the IV of this regime, and will be denoted by $\IV$. Or, put mathematically, the IV is defined~as 
\begin{align}
\label{eq:def:IV}
    \IV \triangleq \sup \left\{ \alpha > 0 \ \middle|\ r = \left\lceil \alpha N \right\rceil, \lim_{N \to \infty} \BLER{N} = 0  \right\} .
\end{align}

\textit{Error Exponent.} 
Define the exponential decay rate---or simply the error exponent (EE)---of the arrive-failure probability \eqref{eq:failure-prob} for $\alpha < \IV$ by (we will prove that the limit exists)
\begin{align}
\label{eq:def:EE}
    \Eblk \triangleq \lim_{N \to \infty} - \frac{1}{N} \log \BLER{N}.
\end{align}

\textit{Goal.} Our goal in this work will be to derive tight bounds on the arrive-failure probability \eqref{eq:failure-prob} and to determine its (exact) EE $\Eblk$ and the corresponding IV $\IV$.

To that end, we will first treat the transmission of a single packet---corresponding to $\lambda = 0$---in \secref{s:1packet}, and then elevate this treatment to $\lambda > 0$ 
in \secref{s:lambda>0}.


\section{Single-Packet Transmission}
\label{s:1packet}

In this section, we treat the case of communicating a single packet 
which is available to the transmitter at time $A = 1$.\footnote{We suppress the link index when $r=1$.}  

Consider first the setting of a single link $r = 1$. The arrive-failure probability \eqref{eq:failure-prob} in this case 
is given by 
\begin{align}
    \BLER{N} 
    = p^N 
    = \Exp{-N \cdot (-\log p)} ,
\end{align}
meaning that the (two-codeword \cite[Ch~5.3]{GallagerBook1968}) EE is $\Eblk = - \log p$ (and meets the sphere-packing bound \cite[Th.~5.8.1]{GallagerBook1968}). 

Consider now the setting of $r$ links. The rest of the section will be devoted to the analysis for the setting of $r$ links. To that end, we will make use of the following definitions.
Denote by $t_i$ 
the time of arrival of the packet at node $i+1$ (over link $i$) for $i \in [r]$, and set $t_0 = 0$.
Denote further the delay caused by link $i \in [r]$ by $\tau_i = t_i - t_{i-1}$.
Clearly, $\{\tau_i | i \in [r] \}$ are independent and geometrically distributed with $\tau_i$ having mean $1/(1-p_i)$.
The arrive-failure probability \eqref{eq:failure-prob} is given, therefore, by 
\begin{align}
\label{eq:delayed:BLER:taus}
    \BLER{N} = \PR{\sum_{i=1}^r \tau_i > N}.
\end{align} 

\begin{remark}
\label{rem:geometric-pdf}
    Unless otherwise stated, we consider the shifted definition of the geometric distribution which counts the number of trials until the first success \textit{including the success}. That is, for a random variable $X$ that is geometrically distributed with success probability $q$, $\PR{X = k} = (1-q)^{k-1} q$ for $k \in \nats$ and zero otherwise.
\end{remark}

We first consider the special case of homogeneous links, 
\begin{align} 
\label{eq:homogeneous}
    p_1 = p_2 = \cdots = p_r \triangleq p
\end{align} 
in \secref{ss:IV:same-channel}, 
and treat the case of heterogeneous links in \secref{ss:IV:different-channels}.

\subsection{Homogeneous Links}
\label{ss:IV:same-channel}

We consider here the case of homogeneous links \eqref{eq:homogeneous}.

\begin{thm}
\label{thm:IV:homogeneous}
    The IV \eqref{eq:def:IV} 
    of transmitting a single packet 
    over independent homogeneous links \eqref{eq:homogeneous} with erasure probability $p$
    is equal to 
    $\IV = 1-p$, 
    and the EE \eqref{eq:def:EE} for $\alpha < \IV$
    is equal to $\Eblk = \KL{\alpha}{1-p}$. 
    Moreover, the arrive-failure 
    probability \eqref{eq:failure-prob} 
    over $r$ links 
    across $N$ time steps
    is bounded as
    \begin{subequations}
    \label{eq:thm:IV:homo:failure-prob}
    \begin{align}
        &\BLER{N} 
        \geq 
        \frac{(1-p) \sqrt{N} \cdot \Exp{- N \cdot \KL{\frac{r - 1}{N}}{1-p} }}{ \sqrt{ 8 (r - 1) (N - r + 1)}} ;
    \label{eq:thm:IV:homo:LB}
     \\ &\BLER{N} \leq 
     \min \bigger\{ \Exp{- (N-1) \cdot \KL{\frac{r}{N - 1}}{1-p} } , 
\nonumber 
    \\* &\ \ \frac{(1-p) \sqrt{N} }{\sqrt{2\pi (r - 1) (N - r + 1)}}
    \cdot \frac{\Exp{- N \cdot \KL{\frac{r - 1}{N}}{1-p} }}{1 - \Exp{- \KL{\frac{r - 1}{N}}{1-p} }} \bigger\} . \quad\ 
    \label{eq:thm:IV:homo:UB:MoT}
    \end{align}
    \end{subequations}
    for $\alpha \triangleq r/N < \IV$, and goes to $1$ for $\alpha > \IV$.
\end{thm}

\begin{IEEEproof}
    We first derive the IV. 
    \begin{subequations}
    \label{eq:proof:IV:homo}
    \begin{align}
        \lim_{N \to \infty} \BLER{N} 
        &= \lim_{N \to \infty} \PR{\frac{1}{r} \sum_{i=1}^r \tau_i > \frac{N}{\left\lceil \alpha N \right\rceil}}
    \label{eq:proof:IV:r=alphaN}
     \\ &= 
         \begin{cases}
            0, & \E{\tau} < \frac{1}{\alpha}
         \\ 1, & \E{\tau} > \frac{1}{\alpha} 
         \end{cases}
         ;
    \label{eq:threshold}
    \end{align}
    \end{subequations}
    where 
    \eqref{eq:proof:IV:r=alphaN} follows form substituting $r = \lceil \alpha N \rceil$ in \eqref{eq:delayed:BLER:taus}, 
    and \eqref{eq:threshold} follows from the (weak) law of large numbers;
    recalling that $\E{\tau} = 1/(1 - p)$ completes the derivation of $\IV$.
    
    We now prove the bounds in \eqref{eq:thm:IV:homo:failure-prob}
    for $\alpha = r/N < 1-p = \IV$.
    We start with proving the first upper bound in \eqref{eq:thm:IV:homo:UB:MoT}:
    \begin{subequations}
    \label{eq:homo:chernoff}
    \begin{align}
        \!\! \BLER{N}
        &\leq \Exp{ - \sup_{\lambda > 0} \big\{ (N-1) \lambda - r \log \E{\Exp{\lambda\tau}} \big\} } 
    \label{eq:homo:chernoff:chernoff}
     \\ &
        \leq \Exp{- \sup_{x > 1} \left\{ \left( N - r - 1 \right) \log x + r \log \frac{1 - p x}{1 - p} \right\} }\quad\ 
    \label{eq:homo:chernoff:MGF}
    \\ & = \Exp{- (N-1) \cdot \KL{\frac{r}{N - 1}}{1-p}}
    \label{eq:homo:chernoff:final}
    \end{align} 
    \end{subequations}
    where
    \eqref{eq:homo:chernoff:chernoff} follows from applying Chernoff's upper bound to \eqref{eq:delayed:BLER:taus}, 
    \eqref{eq:homo:chernoff:MGF} holds by substituting the moment-generating function of a geometric distribution and $x \triangleq \Exp{\lambda}$, 
    and \eqref{eq:homo:chernoff:final} holds for the maximizer 
    $x = \frac{N - r - 1}{p (N-1)}$.
    Moreover, this bound is known to be exponentially tight by Cram\'er's theorem \cite[Ch.~2]{Dembo-Zeitouni:LargeDeviations:Book}, meaning that the EE is $\Eblk = \KL{\alpha}{1-p}$.
    
    We now move on to proving the remaining upper and lower bounds. Since $\{\tau_i\}$ are \iid\ geometric with success probability $1-p$, the arrive-failure probability may be expressed as follows.
    \begin{align}
    \label{eq:Pe:IV:exact}
        \BLER{N} 
        &= \sum_{j=N+1}^\infty {j-1 \choose r-1} (1-p)^r p^{j-r} .
    \end{align}
    Using standard bounds on the binomial coefficient for $n \in \nats$, $k+1 \in [n]$ 
    \cite[Ch.~4.7]{AshBook}: 
    \begin{align} 
        \frac{1}{2} \leq {n \choose k} \Exp{-n \Hb{\frac{k}{n}}} \sqrt{\frac{2k(n-k)}{n}} \leq \sqrt{ \frac{1}{\pi}} ,
    \end{align} 
    we obtain the upper bound
    \begin{align}
        \BLER{N} &\leq (1-p) 
        \sum_{\ell = N}^\infty \frac{\sqrt{\ell} \cdot \Exp{- \ell \cdot \KL{\frac{r - 1}{\ell}}{1-p}}}{\sqrt{2\pi (r - 1) (\ell + 1 - r)}}
\nonumber
     \\  \leq& \frac{(1-p) \sqrt{N} \sum_{\ell = N}^\infty \Exp{- \ell \cdot \KL{\frac{r - 1}{N}}{1-p}}}{\sqrt{2\pi (r - 1) (N + 1 - r )}}
\nonumber
     \\ =& \frac{(1-p) \sqrt{N} }{\sqrt{2\pi (r - 1) (N - r + 1)}}
    \cdot \frac{\Exp{- N \cdot \KL{\frac{r - 1}{N}}{1-p} }}{1 - \Exp{- \KL{\frac{r - 1}{N}}{1-p} }}
\nonumber
\end{align}
    and the lower bound
    \begin{align}
        \BLER{N} &\geq (1-p)
        \sum_{\ell = N}^\infty \frac{\sqrt{\ell} \cdot \Exp{- \ell \cdot \KL{\frac{r - 1}{\ell}}{1-p}}}{ 2 \sqrt{ 2 (r - 1) (\ell+1 - r)}} 
    \nonumber
     \\ & \geq \frac{\sqrt{N} \cdot (1-p) \cdot \Exp{- N \cdot \KL{\frac{r - 1}{N}}{1-p}}}{ \sqrt{ 8 (r - 1) (N - r + 1)}} .
\nonumber
    \end{align}
    Finally, to derive the EE, set $r = \lceil \alpha N \rceil$ in these bounds and take $N$ to infinity to arrive at $\Eblk = \KL{\alpha}{1-p}$.
\end{IEEEproof}

$r$ that grows sublinearly with $N$ results in $\alpha = 0$. Substituting this in \thmref{thm:IV:homogeneous} yields the following immediate result.
\begin{cor}
\label{cor:r=o(N):homogeneous}
    For $r = o(N)$, the EE is equal to that of a single link: $\Eblk = -\log p$. 
\end{cor}


\subsection{Heterogeneous Links}
\label{ss:IV:different-channels}

We now treat the more general case of heterogeneous links.
For simplicity, we concentrate on the case of links having one out of $S \in \nats$ possible (different) erasure probabilities: $P(i) \in [0,1)$ for $i \in [S]$, and denote the \textit{possible erasure-probabilities vector} by
$\bP \triangleq \left( P(1), P(2), \ldots, P(S) \right)$. 

As we show next, the IV and the EE depend on the channel \rm{type}, i.e., on the fraction of channels with a specific erasure probability rather than on the assignment of specific erasure probabilities to specific channel.
Denote by $\bp \triangleq (p_1, p_2, \ldots, p_r)$ the sequence of channel erasure probabilities, 
and by $R(i)$ the number of links with erasure probability $P(i)$ for $i \in [S]$, \ie, the number of components in $\bp$ that are equal to $P(i)$.
Define further the \textit{channels-type} $\bQ_\bp \triangleq \left( Q_\bp(1), Q_\bp(2), \ldots, Q_\bp(S) \right)$ of the sequence $\bp$ via $Q_\bp(i) \triangleq R(i) / r \in \rats \cap [0,1]$. Clearly $\sum_{i=1}^S R(i) = r$, or equivalently, $\bQ_\bp \triangleq \left( Q_\bp(1), Q_\bp(2), \ldots, Q_\bp(S) \right) \in \Delta_{S-1}$, where $\Delta_n$ is the standard $n$-simplex defined in \secref{ss:intro:notation}. 
The set of all channels-types with denominator $r$ and alphabet $[S]$ is denoted by
$\cQ_r \triangleq \left\{ \bQ \in \Delta_{S-1} \middle| r Q(i)+1 \in [r+1], i \in [S] \right\}$. We denote by $T(i)$ a geometrically distributed RV with success probability $1 - P(i)$ for all $i \in [S]$.

We will consider two settings: 
\begin{itemize}
\item 
    A fixed-type setting: $\bQ \in \rats^S \cap \Delta_{S-1}$ is fixed.
\item 
    a probabilistic setting: $\bp$ comprises \iid\ samples according to $\tbQ \in \Delta_{S-1}$.
\end{itemize}

We will concentrate on determining the IV and EE; bounds on the arrive-failure probabilities are derived inside the prove of the remaining theorems in this section but are not explicitly stated in the theorems themselves due to a lack of space.

Consider first the fixed-type setting. 
\begin{thm}
\label{thm:IV:hetero:fixed:delayed}
    The IV of 
    a cascade of links with a fixed channels-type $\bQ \in \rats^S \cap \Delta_{S-1}$ over a possible erasure-probabilities vector $\bP \in \Delta_{S-1}$ equals
    $\IV = \left( \sum_{i=1}^S \frac{Q(i)}{1 - P(i)} \right)^{-1}$. Furthermore, for $r = \left\lceil \alpha N \right\rceil$, the error probability goes to $1$ for $\alpha > \IV$, while for $\alpha < \IV$, the EE $\Eblk^\mathrm{fixed}(\bQ)$ is given by 
    \begin{align}
    \label{eq:fixed-type:EE:via-Chernoff}
        \Eblk^\mathrm{fixed}(\bQ) = (1-\alpha) \log x + \alpha \sum_{i=1}^S Q(i) \log \frac{1 - P(i) x}{1 - P(i)} ,\ 
    \end{align}
    where $x$ is the solution of the equation  
    \begin{align}
    \label{eq:thm:fixed-type:sol}
        \sum_{i=1}^S \frac{Q(i)}{1 - P(i) x} = \frac{1}{\alpha}
    \end{align}
    that lies in the interval $\left( 1, 1/\min (\bP) \right)$.
    Alternatively, the EE may be calculated via the optimization 
    \begin{align}
    \label{eq:fixed-type:EE:via-types}
        \Eblk^\mathrm{fixed}(\bQ) = \min_{U \in \Delta_{S-1}: \atop U(i) \geq \frac{\alpha Q(i)}{1 - P(i)} \forall i \in [S]}  \sum_{i = 1}^S U(i) \KL{\frac{\alpha Q(i)}{U(i)}}{1 - P(i)} . \quad 
    \end{align}
\end{thm}

\begin{IEEEproof}
    Again, we start by deriving the IV. 
    \begin{align}
        \lim_{N \to \infty} \BLER{N} 
        &= \lim_{N \to \infty} \PR{\sum_{i=1}^S \frac{R(i)}{r} \cdot \frac{1}{R(i)} \sum_{\ell:\, p_\ell = P(i)} \tau_{\ell} > \frac{N}{r}}
    \\ &= 
         \begin{cases}
            0, & \sum_{i=1}^S Q(i) \E{T(i)} > \frac{1}{\alpha}
         \\ 1, & \sum_{i=1}^S Q(i) \E{T(i)} < \frac{1}{\alpha}
         \end{cases}
        ;
    \end{align}
    where 
    the first equality follows from \eqref{eq:delayed:BLER:taus} and the definition of $R(i)$ and the law of large numbers, 
    and the second equality follows from the definition of $\bQ$;
    substituting $\E{T(i)} = 1/\left(1 - P(i)\right)$ completes the derivation of~$\IV$.

    We now prove the first characterization of the EE \eqref{eq:fixed-type:EE:via-Chernoff}.\footnote{Assume $N$ is large enough such that $(1 - \alpha) N > 1$.}
    \begin{subequations}
    \label{eq:proof:IV:EE:hetero}
    \begin{align}
       \BLER{N} 
        &\leq \inf_{\lambda > 0} \frac{\prod_{i = 1}^S \left( \E{\Exp{\lambda T(i)}} \right)^{R(i)}}{\Exp{\lambda (N-1)}}
    \label{eq:proof:IV:EE:hetero:Chernoff}
    \\ &\leq \exp \bigg\{ - \sup_{x > 1} \bigg\{ \big( (1-\alpha) N - 1 \big) \log x 
    \\* &\qquad\qquad 
        + r \sum_{i=1}^S Q(i) \log \frac{1 - P(i) x}{1 - P(i)} \bigg\} \bigg\}\ \ 
    \label{eq:proof:IV:EE:hetero:subs}
    \end{align} 
    \end{subequations} 
    where 
    \eqref{eq:proof:IV:EE:hetero:Chernoff} follows from applying Chernoff's upper bound to \eqref{eq:delayed:BLER:taus} and the mutual independence of the RVs in $\{ \tau_\ell \}$,
    and \eqref{eq:proof:IV:EE:hetero:subs} holds since $T(i)$ is geometrically distributed with success probability $1-P(i)$ and by substituting $x = \Exp{\lambda}$. 

    Now, standard calculus shows that the maximizer of the optimization in \eqref{eq:proof:IV:EE:hetero:subs} satisfies 
    $\sum_{i=1}^S \frac{Q(i)}{1 - P(i) x} = \frac{(1-\alpha) N - 1 + r}{r}$.
    Furthermore, this bound is exponentially tight by the G\"artner--Ellis theorem \cite[Ch.~2.3]{Dembo-Zeitouni:LargeDeviations:Book}.
    
    We next prove the alternative characterization of the EE \eqref{eq:fixed-type:EE:via-types} by bounding the arrive-failure probability as follows.\footnote{\label{foot:non-negative-arg-KL}Assume that $N$ is large enough such that the first argument of the KL divergence is lower than or equal to 1.}
    \begin{subequations}
    \label{eq:proof:fixed-type:MoT:UB}
    \begin{align}
       &\BLER{N} 
        \leq \sum_{U \in \cQ_r} \PR{\sum_{\ell: p_\ell = P(i)} \tau_\ell \geq N U(i), \forall i \in [S]}
    \label{eq:proof:fixed-type:MoT:UB:union-bound}
     \\ &= \sum_{U \in \cQ_r} \prod_{i=1}^S \PR{\sum_{\ell: p_\ell = P(i)} \tau_\ell \geq N U(i)}
    \label{eq:proof:fixed-type:MoT:UB:independence}
     \\ &\leq (\alpha N + 2)^S \cdot 
        \exp \Bigg\{ - \!\!\! \min_{U \in \cQ_r: \atop U(i) \geq \frac{\alpha N + 1}{N} \cdot \frac{Q(i)}{1 - P(i)} + \frac{2}{N}}  
        \sum_{i=1}^S \bigg\{ \big( N U(i) - 2 \big)   
    \nonumber
    \\* &\qquad\qquad\qquad\qquad \cdot \KL{\frac{r Q(i)}{N U(i) - 2}}{1-P(i)} \bigg\} \Bigg\}
    \label{eq:proof:fixed-type:MoT:UB:homo-bound}
    \end{align}
    \end{subequations}
    where 
    \eqref{eq:proof:fixed-type:MoT:UB:union-bound} follows from applying the union bound to \eqref{eq:delayed:BLER:taus};
    \eqref{eq:proof:fixed-type:MoT:UB:independence} holds by the independence of $\{\tau_\ell\}$;
    and \eqref{eq:proof:fixed-type:MoT:UB:homo-bound} follows from \thmref{thm:IV:homogeneous}, \footnote{We use the first term in \eqref{eq:thm:IV:homo:UB:MoT} since the second term explodes for $\alpha = 1- p$. That said, since the second term 
    bounds a probability from above, one may always take the minimum between it and and 1 to and use this bound in \eqref{eq:proof:fixed-type:MoT:UB:homo-bound}.}
    and from the bound on the number of types of sequences of a given length \cite[Ch.~11.1]{CoverBook2Edition}.

    Similarly, $\BLER{N}$ may be bounded from below~by 
    \begin{subequations}
    \label{eq:proof:fixed-type:MoT:LB}
    \noeqref{eq:proof:fixed-type:MoT:LB:final} 
    \begin{align}
       &\BLER{N} 
        \geq \max_{U \in \cQ_r} \prod_{i=1}^S \PR{\sum_{\ell: p_\ell = p(i)} \tau_\ell > N U(i)}
    \label{eq:proof:fixed-type:MoT:LB:max-term}
     \\ &\geq \max_{U \in \cQ_r: \atop U(i) \geq \frac{\alpha Q(i)-\frac{1}{N}}{1 - P(i)}} 
        \prod_{i=1}^S \frac{(1 - P(i)) \sqrt{N U(i)}}{\sqrt{8 \left( R(i) - 1 \right) \left( N U(i) + 1 - R(i) \right) }} 
    \nonumber
    \\* &\qquad\qquad\quad \cdot \Exp{- N U(i) \cdot \KL{\frac{R(i) - 1}{N U(i)}}{1-P(i)}} 
    \label{eq:proof:fixed-type:MoT:LB:homo-bound}
     \\ &\geq \frac{ \Exp{- N \!\!\!\!\!\!\!\! \min\limits_{U \in \cQ_r: \atop U(i) \geq \frac{\alpha Q(i)}{1 - P(i)}} \!\! \sum_{i=1}^S U(i) \KL{\frac{\alpha Q(i) - \frac{1}{N}}{U(i)}}{1-P(i)}} }{\left(1 - \max (\bP) \right)^{-S} \cdot (N+1)^S} \quad\ \ 
    \label{eq:proof:fixed-type:MoT:LB:final}
    \end{align}
    \end{subequations}
    where 
    \eqref{eq:proof:fixed-type:MoT:LB:max-term} follows from \eqref{eq:delayed:BLER:taus} and the independence of $\{\tau_\ell\}$, 
    and \eqref{eq:proof:fixed-type:MoT:LB:homo-bound} follows from 
    \eqref{eq:thm:IV:homo:LB}.  
    
    Since $\bigcup_{r = 1}^\infty \cQ_r = \rats \cap [0,1]$ is dense in $[0,1]$, by taking the limit $N \to \infty$, we attain \eqref{eq:fixed-type:EE:via-types}. 
\end{IEEEproof}

\begin{cor}
\label{cor:r=o(N):fixed-type}
    For $r = o(N)$, the EE is equal to that of the worst link: $\Eblk = -\log \min (\bP)$.
\end{cor}

\begin{remark}
\label{eq:KKT:fixed-type}
    We have derived two different expressions for the EE in \eqref{eq:fixed-type:EE:via-Chernoff}
    and \eqref{eq:thm:fixed-type:sol}, and have proved implicitly that they are equal. 
    Using the Karush--Khun--Tucker conditions and convexity arguments \cite[Ch.~5.5]{BoydBook} one may verify explicitly that the solution of \eqref{eq:thm:fixed-type:sol} yields \eqref{eq:fixed-type:EE:via-Chernoff}.
\end{remark}

For the probabilistic setting, the following result holds.

\begin{thm}
\label{thm:IV:hetero:prob:delayed}
    The IV of a cascade of links with \iid\ erasure probabilities according to $\tbQ \in \Delta_{S-1}$ over a possible erasure-probabilities vector $\bP \in \Delta_{S-1}$
    equals
    $\IV = \left( \sum_{i=1}^S \frac{\tQ(i)}{1 - P(i)} \right)^{-1}$. 
    Furthermore, for $r = \left\lceil \alpha N \right\rceil$, the error probability goes to $1$ for $\alpha > \IV$, while for $\alpha < \IV$, the EE $\Eblk^\mathrm{prob}(\tbQ)$ is given by
    \begin{align}
    \label{eq:thm:EE:prob:Chernoff}
        \Eblk^\mathrm{prob}(\tbQ) = (1-\alpha) \log x - \alpha \log \left( \sum_{i=1}^S \tQ(i) \frac{1 - P(i)}{1 - P(i) x} \right) ,\quad\ 
    \end{align}
    where $x$ is the solution of the equation 
    \begin{align}
        \sum_{i=1}^S \tQ(i) \cdot \left\{ 1 - P(i) \right\} \cdot \frac{1 - \alpha - P(i) x}{\left( 1 - P(i) x \right)^2} = 0
    \end{align}
    that lies in the interval $x \in \left( 1, 1/\min (P) \right)$.
    Alternatively, the EE may be calculated via the optimization 
    \begin{align}
    \label{eq:thm:EE:prob:MoT}
        \Eblk^\mathrm{prob}(\tbQ) = \min_{\bQ \in \Delta_{S-1}} \left\{ \Eblk^\mathrm{fixed}(\bQ) + \alpha \KL{\bQ}{\tbQ} \right\} ,
    \end{align}
    where $\Eblk^\mathrm{fixed}$ is given in \thmref{thm:IV:hetero:fixed:delayed}.
\end{thm}

\begin{IEEEproof}
    Since $\{\tau_i | i \in [r]\}$ are \iid\ Geometric mixtures: 
    \begin{align}
        \PR{\tau = \ell} = \sum_{i = 1}^S \tQ(i) P(i)^{\ell-1} (1 - P(i)) ,
    \end{align}
    $\IV$ may be derived as in \eqref{eq:proof:IV:homo}
    by plugging in 
    \begin{align}
        \E{\tau} = \sum_{i = 1}^S \frac{\tQ(i)}{1 - P(i)}.
    \end{align}
    
    Moreover, the first characterization of the EE \eqref{eq:thm:EE:prob:Chernoff} 
    can be derived as in the first derivation of the EE in the proof of \thmref{thm:IV:homogeneous}: 
    Since the error probability is given by \eqref{eq:delayed:BLER:taus}, 
    the optimal EE according to Cram\'er's theorem \cite[Ch.~2]{Dembo-Zeitouni:LargeDeviations:Book} is given by
    \begin{align}
        \Eblk^\mathrm{prob}(\tbQ) = \sup_{\lambda > 0} \left\{ \lambda - \alpha \log \E{\Exp{\lambda \tau}} \right\} , 
    \end{align}
    which can be shown to equal to \eqref{eq:thm:EE:prob:Chernoff} by standard calculus.
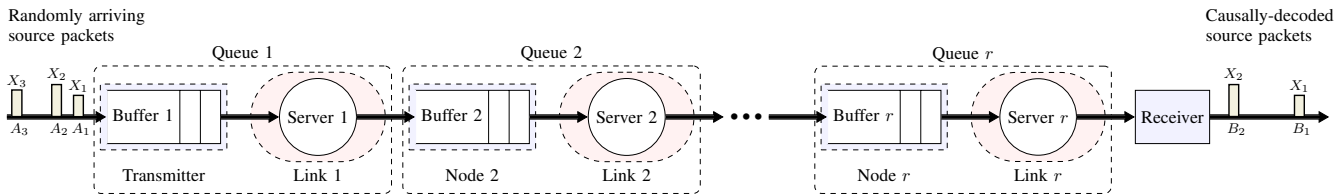
\begin{figure*}[t]
    \resizebox{\textwidth}{!}{\begin{tikzpicture}[auto, arrow/.style={very thick, ->, >=stealth'},start chain=going right,>=latex,node distance=.13\columnwidth,>=latex']
    \node[coordinate] (input) {};
    \node[FIFO, right of = input, node distance = .35\columnwidth, fill=white] (fifo1) {Buffer 1};
    \draw[blue!5,very thick] (fifo1.south west) -- (fifo1.north west);
    \begin{pgfonlayer}{background}
        \node[fill = blue!5, draw, dashed, fit = (fifo1)] (enc) {};
    \end{pgfonlayer}
    \node[below of = enc] (Tx-text) {Transmitter};

    \node[draw,circle, right = of fifo1, fill=white] (server1) {Server 1};
    \begin{pgfonlayer}{background}
        \node[fill = red!5, draw, dashed, rounded rectangle, fit = (server1)] (link1) {};
    \end{pgfonlayer}
    \node[below of = link1] (link1-text) {Link 1};

    \node[FIFO, right = of server1, node distance = .35\columnwidth, fill=white] (fifo2) {Buffer 2};
    \draw[blue!5,very thick] (fifo2.south west) -- (fifo2.north west);
    \begin{pgfonlayer}{background}
        \node[fill = blue!5, draw, dashed, fit = (fifo2)] (node2) {};
    \end{pgfonlayer}
    \node[below of = node2] (node2-text) {Node 2};

    \node[draw,circle, right = of fifo2, fill=white] (server2) {Server 2};
    \begin{pgfonlayer}{background}
        \node[fill = red!5, draw, dashed, rounded rectangle, fit = (server2)] (link2) {};
    \end{pgfonlayer}
    \node[below of = link2] (link2-text) {Link 2};
    
    \node[mwe, right = of server2] (etc) {$\bullet\bullet\bullet$};

    \node[FIFO, right = of etc, node distance = .35\columnwidth, fill=white] (fifoR) {Buffer $r$};
    \draw[blue!5,very thick] (fifoR.south west) -- (fifoR.north west);
    \begin{pgfonlayer}{background}
        \node[fill = blue!5, draw, dashed, fit = (fifoR)] (nodeR) {};
    \end{pgfonlayer}
    \node[below of = nodeR] (nodeR-text) {Node $r$};

    \node[draw,circle, right = of fifoR, fill=white] (serverR) {Server $r$};
    \begin{pgfonlayer}{background}
        \node[fill = red!5, draw, dashed, rounded rectangle, fit = (serverR)] (linkR) {};
    \end{pgfonlayer}
    \node[below of = linkR] (linkR-text) {Link $r$};
    
    \node[block, right = of serverR] (dec) {Receiver};
    \node[coordinate, right of = dec, node distance = .35\columnwidth] (output) {};

    \begin{pgfonlayer}{background}
        \node[draw, rounded corners, dashed, fit = (enc) (link1) (Tx-text) (link1-text)] (Q1) {};
    \end{pgfonlayer}
    \node[above of = Q1, node distance = 15mm] {Queue 1};

    \begin{pgfonlayer}{background}
        \node[draw, rounded corners, dashed, fit = (node2) (link2) (node2-text) (link2-text)] (Q2) {};
    \end{pgfonlayer}
    \node[above of = Q2, node distance = 15mm] {Queue 2};

    \begin{pgfonlayer}{background}
        \node[draw, rounded corners, dashed, fit = (nodeR) (linkR) (nodeR-text) (linkR-text)] (Qr) {};
    \end{pgfonlayer}
    \node[above of = Qr, node distance = 15mm] {Queue $r$};

    \draw[BitPipe] (input) -- node [tap, pos=0.1, scale=0.5, minimum width= .4cm, above] {}  node [pos=0.1, below] {\footnotesize{ $A_3$}} node[pos=.1, above, minimum height = 14mm] {\footnotesize $X_3$}  
    node [tap, pos=0.5, scale = 0.6, minimum width = 0.32cm, above] {} node [pos=0.5, below] {\footnotesize{ $A_2$}} node[pos=.5, above, minimum height = 16.5mm] {\footnotesize $X_2$}
    node [tap, pos=0.72, scale=0.4, minimum width = .485cm, above] {} node [pos=0.72, below] {\footnotesize{ $A_1$}} node[pos=.72, above, minimum height = 12mm] {\footnotesize $X_1$}
    node[above, pos=.7]{\begin{tabular}{l}
          Randomly arriving
       \\ source packets
          \\ \\ \\ \\
     \end{tabular}}
    (fifo1);

    \draw[BitPipe] (dec) -- 
    node [tap, pos=0.2, scale=0.6, minimum width= .32cm, above] {}  node [pos=0.2, below] {\footnotesize{ $B_2$}}
    node[pos=.2, above, minimum height = 16.5mm]{\footnotesize $X_2$} 
    node [tap, pos=0.75, scale=0.4, minimum width = .485cm, above] {} node [pos=0.75, below] {\footnotesize{ $B_1$}} node[pos=.75, above, minimum height = 12mm] {\footnotesize $X_1$}
    node [above, pos=0.5] {\begin{tabular}{l} 
                                Causally-decoded 
                             \\ source packets 
                             \\ \\ \\ \\
                           \end{tabular}} 
                           (output);

    \draw[BitPipe] (fifo1) -- (server1);
    \draw[BitPipe] (server1) -- (fifo2);
    \draw[BitPipe] (fifo2) -- (server2);
    \draw[BitPipe] (server2) -- (etc);
    \draw[BitPipe] (etc) -- (fifoR);
    \draw[BitPipe] (fifoR) -- (serverR);
    \draw[BitPipe] (serverR) -- (dec);
\end{tikzpicture}}
    \caption{Block diagram of the system model: Queueing theory view.} 
    \label{fig:queues}
\end{figure*}

    To derive the the second characterization of the EE \eqref{eq:thm:EE:prob:MoT}, we bound the error probability from above as follows.
    \begin{subequations}
    \label{eq:proof:hetero:prob:MoT}
    \noeqref{eq:proof:hetero:prob:MoT:UB:smooth}
    \begin{align}
        &\BLER{N} 
        = \sum_{\bQ \in \cQ_r} \CPR{\sum_{i=1}^r \tau_i > N}{\bQ_p = \bQ} \PR{\bQ_p = \bQ} \ 
    \label{eq:proof:hetero:prob:MoT:UB:smooth}
     \\ &\leq \sum_{\bQ \in \cQ_r} \!\! \Exp{-N \left( \Eblk^\mathrm{fixed}(\bQ) + o(1) \right)} \Exp{-r \KL{\bQ}{\tbQ}} \quad 
    \label{eq:proof:hetero:prob:MoT:UB:TypeProb}
     \\ &\leq (r+1)^S \Exp{- N \min_{\bQ \in \cQ_r} \left\{ \Eblk^\mathrm{fixed}(\bQ) + \alpha \KL{\bQ}{\tbQ} + o(1) \right\}} \!,
    \label{eq:proof:hetero:prob:MoT:UB:type-num}
    \end{align}
    \end{subequations}
    where \eqref{eq:proof:hetero:prob:MoT:UB:smooth} follows from applying the law of total probability to \eqref{eq:delayed:BLER:taus}, \eqref{eq:proof:hetero:prob:MoT:UB:TypeProb} follows from \thmref{thm:IV:hetero:fixed:delayed} (and its proof) and \cite[Thm.~11.1.4]{CoverBook2Edition}, 
    and \eqref{eq:proof:hetero:prob:MoT:UB:type-num} follows from 
    \cite[Thm~11.1.1]{CoverBook2Edition}. 
    
    Similarly, we bound the error probability from below by 
    \begin{align}
        &\BLER{N} 
        = \sum_{\bQ \in \cQ_r} \CPR{\sum_{i=1}^r \tau_i > N}{\bQ_p = \bQ} \PR{\bQ_p = \bQ} 
    \nonumber
     \\&\geq \frac{ \Exp{- N \min_{\bQ \in \cQ_r} \left\{ \Eblk^\mathrm{fixed}(\bQ) + \alpha + o(1) + \KL{\bQ}{\tbQ} \right\}} }{(r+1)^S}
    \end{align}
    where the inequality follows from \thmref{thm:IV:hetero:fixed:delayed} (and its proof) and \cite[Thm.~11.1.4]{CoverBook2Edition}. 

    Since $\bigcup_{r = 1}^\infty \cQ_r = \rats \cap [0,1]$ is dense in $[0,1]$, by taking the limit $N \to \infty$, we attain \eqref{eq:thm:EE:prob:MoT}.
\end{IEEEproof}

\begin{remark}
\label{eq:KKT:probabilistic}
    Similarly to \remref{eq:KKT:fixed-type}, we have derived two different expressions for the EE in 
    \eqref{eq:thm:EE:prob:Chernoff} and 
    \eqref{eq:thm:EE:prob:MoT}, and have proved implicitly that they are equal. 
    Using the Karush--Khun--Tucker conditions and convexity arguments \cite[Ch.~5.5]{BoydBook} one may verify explicitly that the solution of \eqref{eq:thm:EE:prob:MoT} yields \eqref{eq:thm:EE:prob:Chernoff}.
\end{remark}

For a comparison of the EEs in the fixed-type and probabilistic settings is available in \figref{fig:EEs} in \secref{ss:extensions:1packet:instant-hops}.


\section{Positive Arrival Rate} 
\label{s:lambda>0}

In this section, we treat the general setting of $\lambda > 0$.

Clearly, since the queueing time of the first packet at each node---the time it waits before the node attempts to send it over its link (before it is served)---is zero, 
the analysis and results of \secref{s:1packet} remain valid for this packet. 

However, for each subsequent packet, the (mean) queueing time increases due to earlier packets that have not been yet served.\footnote{For analysis of the transient behavior, see \cite{Mohanty-Panny:Geo/Geo/1:transient:geometric-approach,Kim:Geo/Geo/1:transient}.}
That said, if 
\begin{align} 
\label{eq:lambda<mu}
    \lambda &< 1 - p_i, & \forall i \in [r]
\end{align} 
the system is stable and reaches a steady state behavior (see, \eg, \cite{Hsu-Burke:Burke-Theorem:Geo/Geo/1:COM1976}); if \eqref{eq:lambda<mu} is not satisfied for some $i$, then that queue will grow indefinitely long, as will also the delay.
We therefore concentrate on analyzing the system in steady state assuming  \eqref{eq:lambda<mu} holds.

To that end, we appeal to results from queueing theory that allow elevating the results of \secref{s:1packet} to the setting of $\lambda > 0$ when the system reaches steady state.

Indeed the system model of \secref{s:model} may be viewed as a queueing scenario of a cascade of queues, as depicted in \figref{fig:queues}: 
packets that arrive to a node before previous packets have been successfully sent 
are queued in a buffer and are sent in their order or arrival over the link. 
Since the successful transmission events over each link are \iid\ Bernoulli, 
each link may be viewed as a server with \iid\ geometric service times. The queues are further independent, meaning that the service times of different queues are mutually independent. 

We will consider an input arrival process with \iid\ geometric interarrival times in \secref{ss:lambda>0:Bernoulli}, and more general arrival processes in \secref{ss:lambda>0:non-Bernoulli}.

\begin{figure*}[t]
  \begin{minipage}[t]{0.31\textwidth}
    \includegraphics[width=1.1\textwidth,height=40mm]{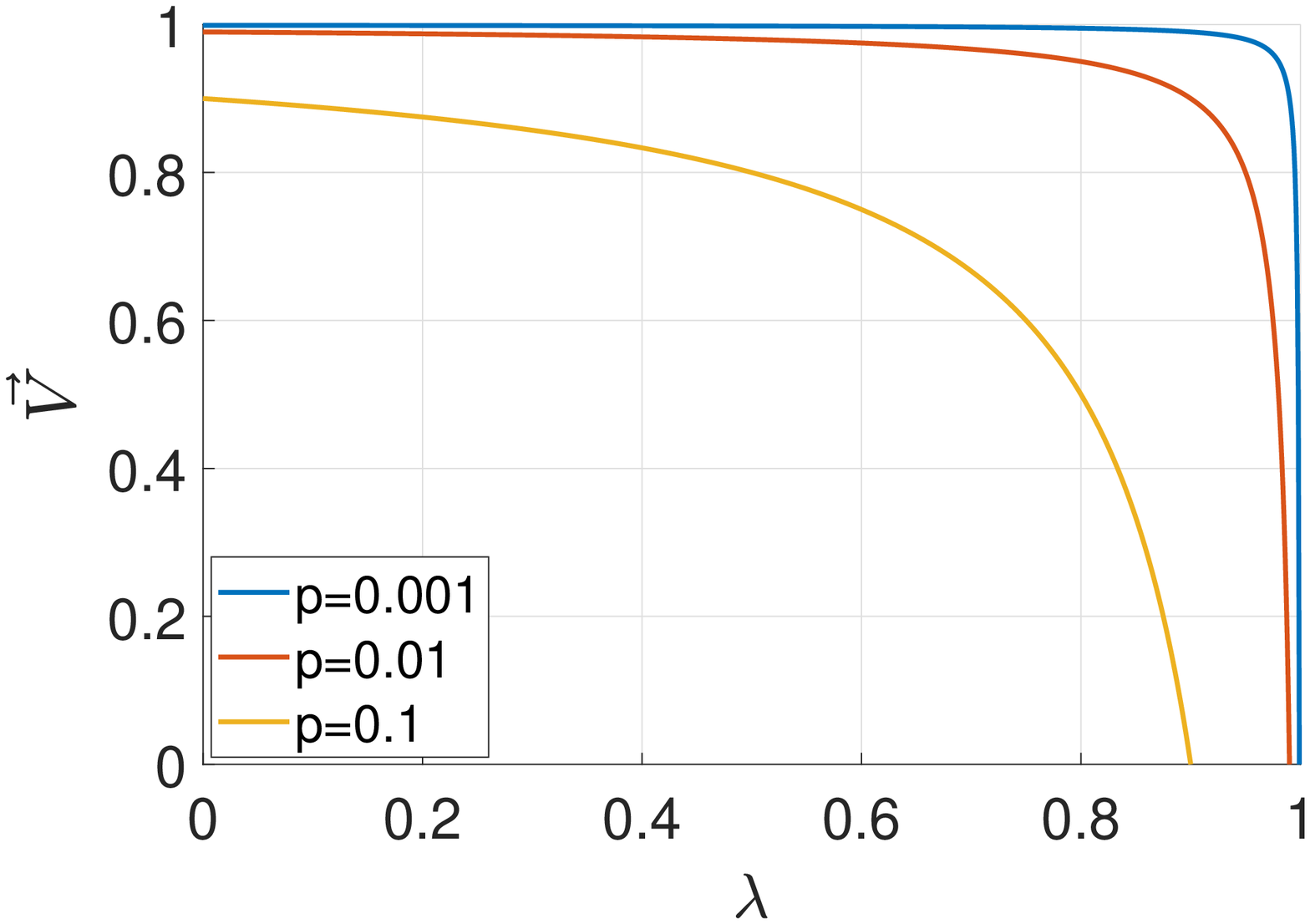}
    \caption{The IV as a function of the arrival rate $\lambda$ for homogeneous links with $p = 0.001, 0.01, 0.1$.}
    \label{fig:IV-vs-lambda}
  \end{minipage}
\ \ \ 
  \begin{minipage}[t]{0.31\textwidth}
    \includegraphics[width=1.05\textwidth,height=40mm]{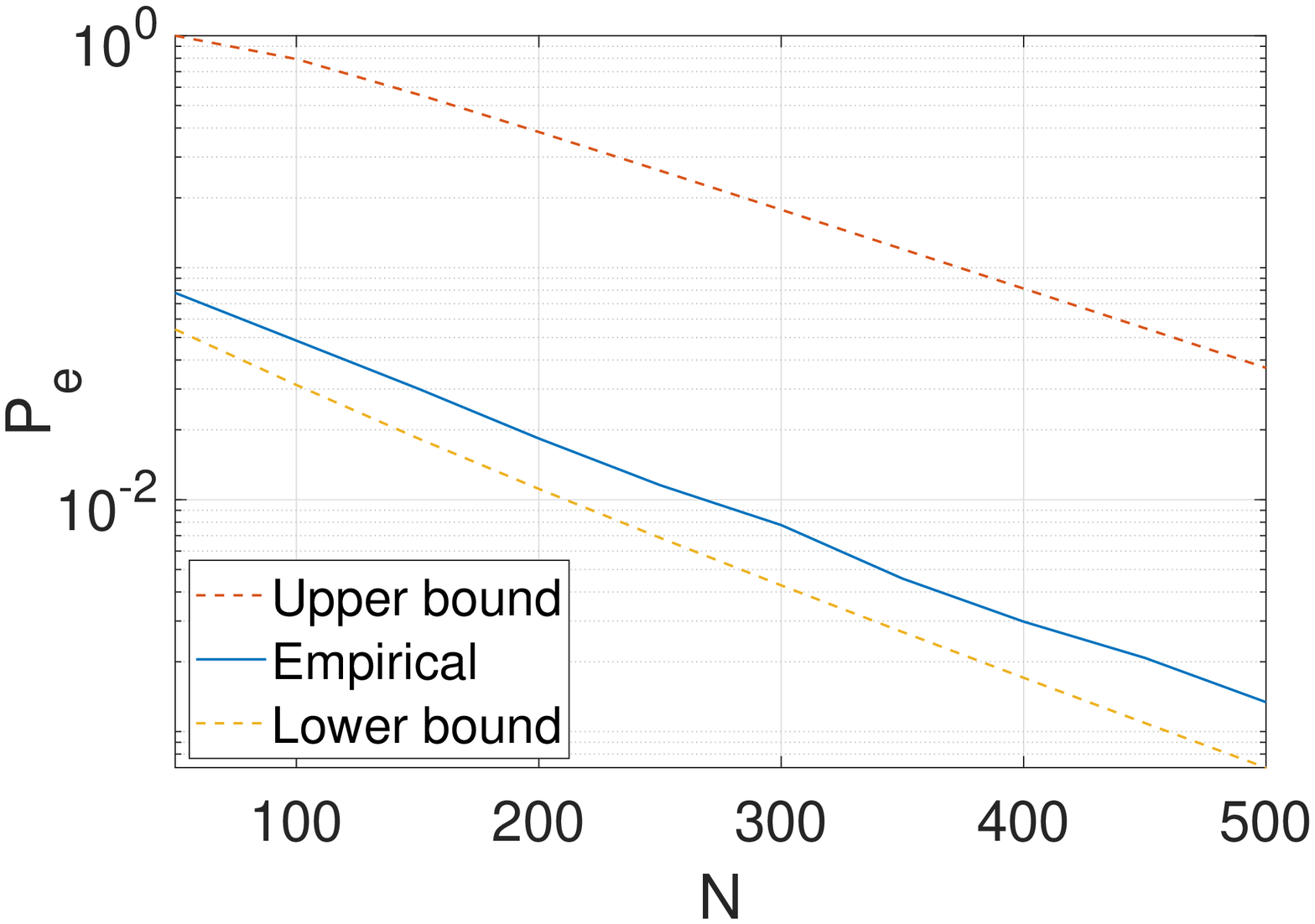}
    \caption{Empirical, upper bound and lower bound (of \thmref{thm:IV:homogeneous} with the extension of \corref{cor:lambda>0:all-results}) $P_e$ curves versus the allowed E2E delay $N$ for \iid\ geometric interarrival times with $\lambda=0.5$ and $p=0.01$, for $\alpha = 0.96$.}
    \label{fig:upper-lower-hom-diff-arrival-vs-N}
  \end{minipage}
\ \ \ 
  \begin{minipage}[t]{0.31\textwidth}
    \includegraphics[width=1.05\textwidth,height=40mm]{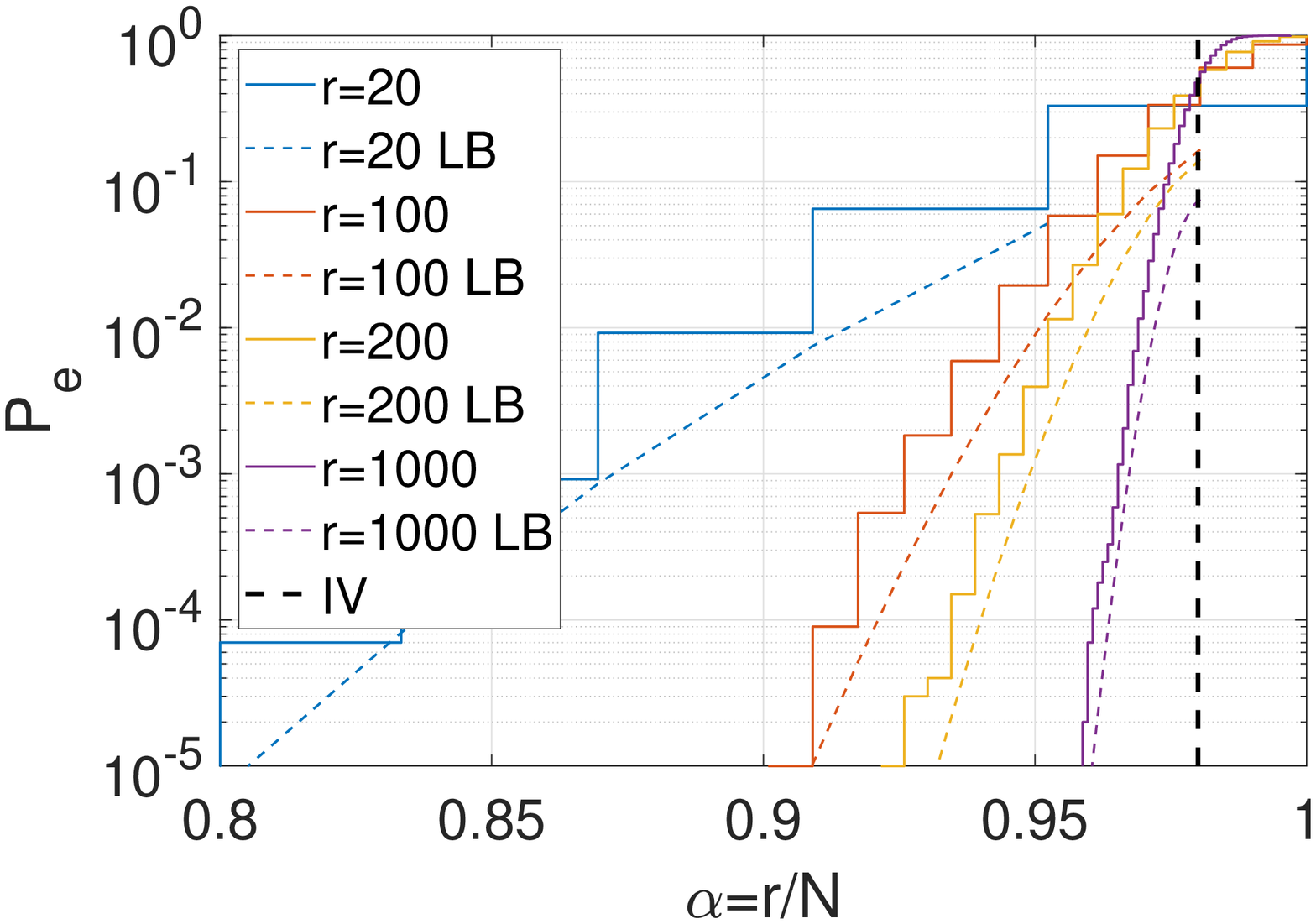}
    \caption{Empirical and lower bound (of \thmref{thm:IV:homogeneous} with the extension of \corref{cor:lambda>0:all-results}) $P_e$ curves as a function of $\alpha$ for \iid\ geometric interarrival times with $\lambda=0.5$ and $p=0.01$,
    with $r = 20, 100, 200, 1000$ relays.}
    \label{fig:Hom-Arrival-Geo-vs-r}
  \end{minipage}
\end{figure*}

\subsection{IID Geometric Interarrival Times}
\label{ss:lambda>0:Bernoulli}

We will adopt Kendall's notation \cite[Ch.~10]{Switching-Traffic-Networks:Book:Hui}. Specifically, we will denote by Geo/Geo/1 queues with an arrival process with \iid\ geometric interarrival times, and \iid\ geometric service times.

Since in our model, the input process has \iid\ interarrival times, the first queue is a Geo/Geo/1 queue.

The following theorem is the discrete-time counterpart of the renowned Burke theorem \cite{Burke:M/M/1:1956,Reich:Tandem:M/M/1:sojourn-times:1957} for Geo/Geo/1 queues, which is due to Hsu and Burke \cite{Hsu-Burke:Burke-Theorem:Geo/Geo/1:COM1976} (see also \cite[Ch.~11.1]{Switching-Traffic-Networks:Book:Hui}, \cite{Prabhakar-Gallager:discrete-time-Burke:IT2003}); the distribution of the waiting time---the elapsed time between arrival to and departure from the queue---is due to \cite{Pujolle-Claude-Seret:Geo/Geo/1:sojourn-time:1986}, \cite[Cor.~2.2]{Desert-Daduna:Geo/Geo/1:sojourn-time:2002}.\footnote{We follow the convention that the waiting time includes both the queueing time and the service time.}
\begin{thm}[\!\!{\cite{Hsu-Burke:Burke-Theorem:Geo/Geo/1:COM1976}, \cite{Pujolle-Claude-Seret:Geo/Geo/1:sojourn-time:1986}, \cite[Cor.~2.2]{Desert-Daduna:Geo/Geo/1:sojourn-time:2002}}]
\label{thm:Burke}
    Assume a Geo/Geo/1 queue with arrival rate $\lambda$ in steady state and geometric service time with success probability $1-p$. Then,\footnote{Burke's original paper \cite{Burke:M/M/1:1956} states only the first two properties. All the three properties are proved by Reich \cite{Reich:Tandem:M/M/1:sojourn-times:1957}. We follow the exposition for both M/M/1 and Geo/Geo/1 queues of Hui \cite{Switching-Traffic-Networks:Book:Hui}.}
    \begin{enumerate}
    \item 
    \label{prop:Burke:departure-process}
        The interarrival times of the departure process are \iid\ geometric with mean $1/\lambda$.
    \item 
    \label{prop:Burke:num_customers}
        The number of packets in the queue at time $t$ is independent of the departure process prior to time $t$.
    \item 
    \label{prop:Burke:packet-departure-independent}
        For a particular packet, the waiting time is independent of the departure process before its departure and is geometrically distributed with success probability $1-\frac{p}{1-\lambda}$.
    \end{enumerate}
\end{thm} 

By \propref{prop:Burke:departure-process} of \thmref{thm:Burke}, for our cascade of stable independent queues (see \figref{fig:queues}) in steady state, the arrival process to each of them, being the departure process of the previous queue, has \iid\ geometric interarrival times with mean $1/\lambda$, \ie, all are Geo/Geo/1 queues. Moreover,  by \propref{prop:Burke:packet-departure-independent}, the waiting time of a particular packet in queue $i$ is independent of the arrival process to queue $i$ (and therefore also to subsequent queues) prior to the packet's departure of queue $i$. Therefore, 
the waiting times at the different queues are mutually independent. 
The latter result, stated formally next, was first proved by Reich \cite{Reich:Tandem:M/M/1:sojourn-times:1957,Reich:Tandem:M/M/1:IID-system-times:1963} for the continuous-time variant of the problem (see also \cite[Ch.~11.1]{Switching-Traffic-Networks:Book:Hui}). A similar argument may be applied with respect to \propref{prop:Burke:num_customers}.
\begin{thm}[\!\!\!{\cite[Ch.~11.1]{Switching-Traffic-Networks:Book:Hui}, \cite{Prabhakar-Gallager:discrete-time-Burke:IT2003}}]
\label{thm:queue-independence}
    Assume a cascade of $r$ independent queues in steady state, with queue $i \in [r]$ having a geometric service times of means $1/(1-p_i)$, satisfying \eqref{eq:lambda<mu}. Assume further an input process with \iid\ geometric interarrival times of mean $1/\lambda$. Then,
    \begin{enumerate}
    \item 
        Each of the queues is Geo/Geo/1 with input arrival rate~$\lambda$.
    \item 
        The number of packets in each of the queues at a given time are mutually independent.
    \item 
        The waiting times of a packet in the different queues are mutually independent and are geometrically distributed with success probabilities $1 - \frac{p_i}{1 - \lambda}$ for $i \in [r]$.
    \end{enumerate}
\end{thm} 

Ths.~\ref{thm:Burke} and \ref{thm:queue-independence} suggest that in our model of interest of \secref{s:model}, 
that each packet in steady state experiences the same delays as if it were the only packet sent but with erasure probabilities $\{p_i/(1-\lambda) | i \in [r]\}$ in lieu of $\{p_i | i \in [r]\}$. This is formally state next.
\begin{cor}
\label{cor:lambda>0:all-results}
    Consider the model of \secref{s:model} with \iid\ geometric interarrivals with success (arrival) probability $\lambda$ in steady state.\footnote{\label{foot:steady-state} This means in turn that \eqref{eq:lambda<mu} is satisfied for each link individually, \ie, for $p$ in the homogeneous-links case \eqref{eq:homogeneous}, and for $P(i)$ in the heterogeneous-links case for all $i \in [S]$.}
    Then, \thmref{thm:IV:homogeneous} (and in \corref{cor:r=o(N):homogeneous}) hold with $p$ replaced by $p/(1-\lambda)$, and Ths.~\ref{thm:IV:hetero:fixed:delayed} and \ref{thm:IV:hetero:prob:delayed} (and \corref{cor:r=o(N):fixed-type}) hold 
    with $P(i)$ replaced by $P(i)/(1 - \lambda)$ for all $i \in [S]$.
\end{cor}

\subsection{Stationary Ergodic \& Deterministic Arrival Processes}
\label{ss:lambda>0:non-Bernoulli}

Up until now we have concentrated on the setting of \iid\ geometric interarrival times. 
We now move to treating more general arrival processes. To that end, we use a result of Mountford and Prabhakar \cite{Mountford-Prabhakar:M/M/1-queues:convergence:1995} for continuous-time processes that may be adopted also to discrete-time processes \cite[Sec.~II]{Prabhakar-Gallager:discrete-time-Burke:IT2003}.
\begin{thm}
\label{thm:inf-num-iid-queues:departures-convergence}
    Assume a cascade of $r$ \iid\ queues in steady state with geometric service times with mean $1/(1-p)$ satisfying \eqref{eq:lambda<mu}. 
    Assume further a stationary ergodic arrival-time process of mean $1/\lambda$.
    Then, in the limit of $r \to \infty$, 
    the departure process of the last queue 
    has \iid\ geometric interarrival times of mean $1/\lambda$.
\end{thm}

The following corollary extends the result of the last theorem to 
independent queues with different service rates;
its proof is a straightforward adaptation of the proof for \iid\ queues and is therefore omitted.
\begin{cor}
\label{cor:inf-num-independent-queues:departures-convergence}
    Assume a cascade of $r$ independent queues in steady state, with queue $i \in [r]$ having a geometric service time with mean $1/(1-p_i)$, satisfying \eqref{eq:lambda<mu}. 
    Assume further a stationary ergodic arrival-time process of mean $1/\lambda$.
    Then, in the limit of $r \to \infty$, 
    the departure process of the last queue 
    has \iid\ geometric interarrival times with mean $1/\lambda$.
\end{cor}

Although a process with deterministic interarrival times is not stationary ergodic, the results of  \thmref{thm:inf-num-iid-queues:departures-convergence} and \corref{cor:inf-num-independent-queues:departures-convergence} readily apply to such processes as well. 
Again, since the proof for deterministic interarrivals is a simple adaptation of the proof of \thmref{thm:inf-num-iid-queues:departures-convergence}, we omit it in the interest of space.
\begin{cor}
\label{cor:inf-num-independent-queues:departures-convergence:deterministic}
    Assume a cascade of $r$ independent queues in steady state, with queue $i \in [r]$ having a geometric service time with mean $1/(1-p_i)$, satisfying \eqref{eq:lambda<mu}. 
    Assume further a process with deterministic interarrival times $1/\lambda \in \nats$.
    Then, in the limit of $r \to \infty$, 
    the departure process of the last queue 
    has \iid\ geometric interarrival times of mean $1/\lambda$.
\end{cor}

Cors.~\ref{cor:inf-num-independent-queues:departures-convergence} and \ref{cor:inf-num-independent-queues:departures-convergence:deterministic} 
imply that the IV and the EEs of \corref{cor:lambda>0:all-results} remain the same for any stationary ergodic interarrival process having the same average arrival rate $\lambda$ as well as for processes having deterministic interarrival times.
\begin{cor}
\label{cor:lambda>0:all-results:non-geometric}
    Consider the model of \secref{s:model} with a stationary ergodic arrival process of arrival rate $\lambda$ or with a process with deterministic interarrival times $1/\lambda \in \nats$, in steady state.\footnote{The result holds also for a deterministic arrival process with arrivals at $\lfloor i \cdot a \rfloor$ times for a fixed $a \in \rats$, $a > 1$ and all $i \in \ints$.}
    Then, the expressions for the IV and the EE of \thmref{thm:IV:homogeneous} (and \corref{cor:r=o(N):homogeneous}) 
    hold with $p$ replaced by $p/(1-\lambda)$, 
    and Ths.~\ref{thm:IV:hetero:fixed:delayed} and \ref{thm:IV:hetero:prob:delayed} (and \corref{cor:r=o(N):fixed-type}) hold 
    with $P(i)$ replaced by $P(i)/(1 - \lambda)$ for all $i \in [S]$.
\end{cor}


\section{Numerical Results}
\label{s:numeric}

Consider first the homogeneous-links setting \eqref{eq:homogeneous}. 
\figref{fig:IV-vs-lambda} depicts the IV for a stream of packets versus the arrival rate $\lambda$ for several values~of~$p$. Interestingly, this figure shows that for reasonable erasure probabilities, in case of a low IV, a modest reduction in the arrival rate increases the IV substantially, which is explained by the $\IV$ being a shifted negative-sign hyperbolic function of $(1-\lambda)$.

\begin{figure*}
    \begin{subfigure}[t]{.33\textwidth}
        \centering
	    \includegraphics[width=1.05\textwidth]{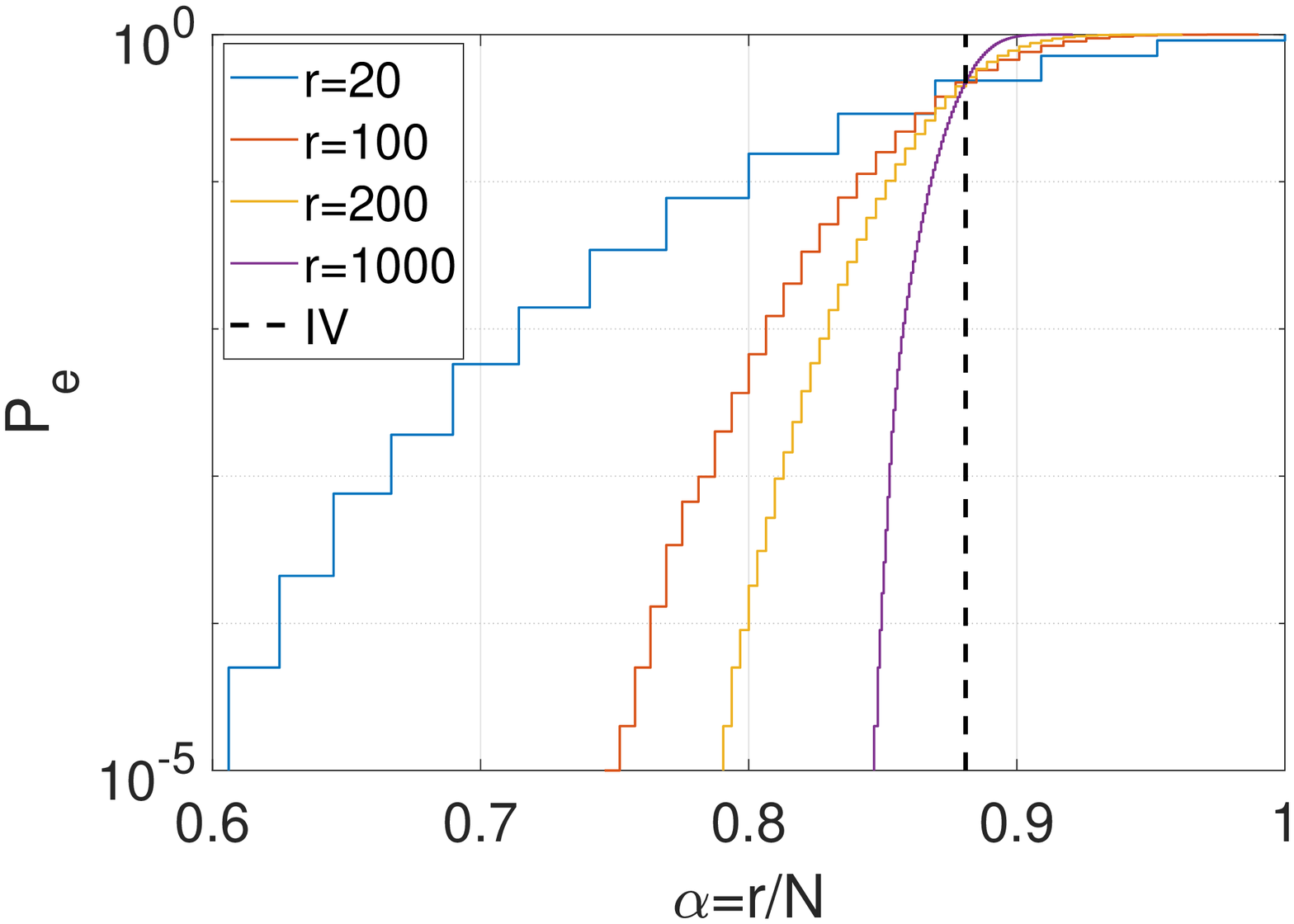}
        \caption{Geometric \iid\ interarrival times.}
    \label{fig:Hetero:Geo-vs-r}
    \end{subfigure}
    \begin{subfigure}[t]{.33\textwidth}
        \centering
        \includegraphics[width=1.05\textwidth]{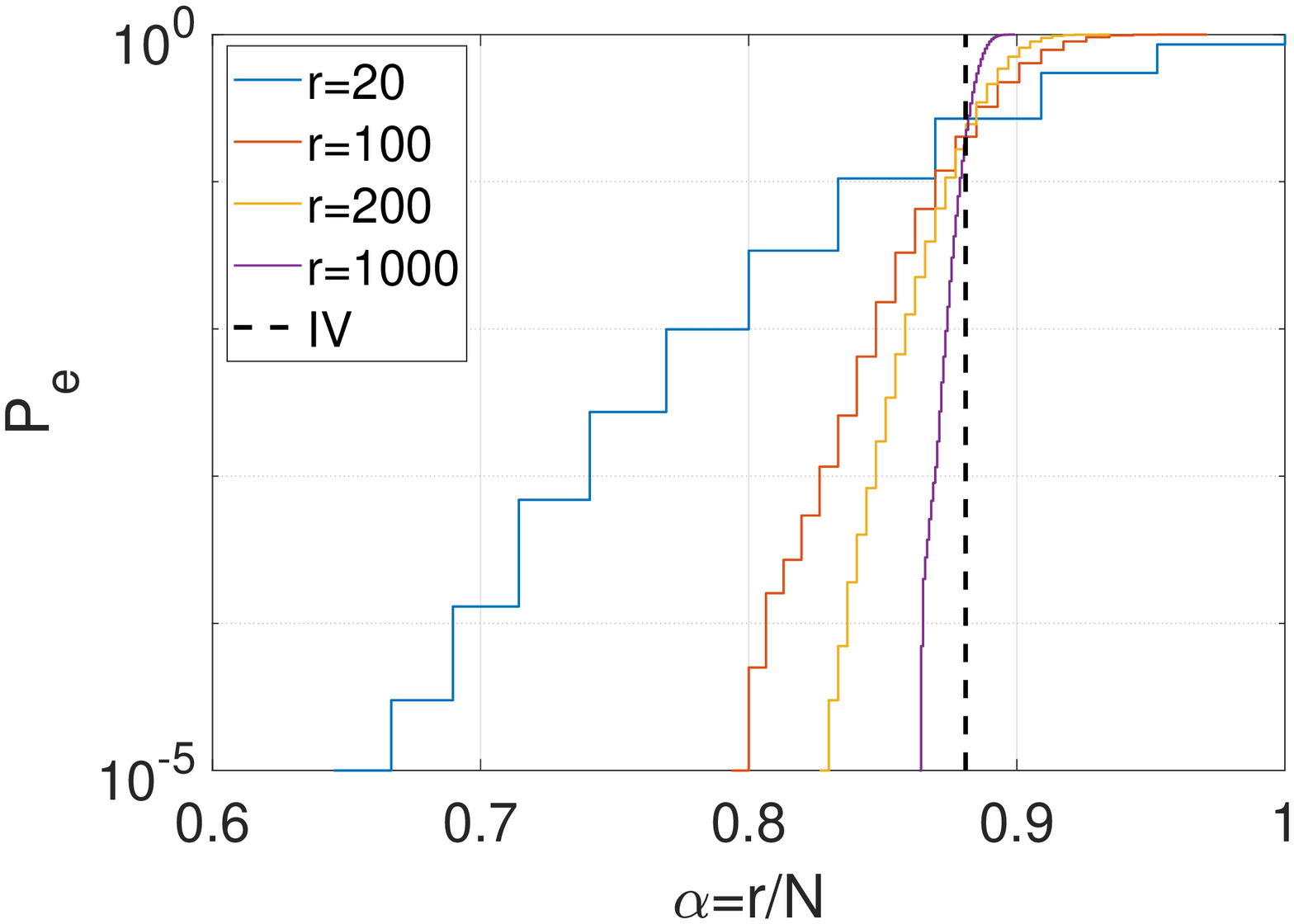}
        \caption{Deterministic interarrival times.}
    \label{fig:Hetero:Det-vs-r}
    \end{subfigure}
    \begin{subfigure}[t]{.33\textwidth}
        \centering
        \includegraphics[width=1.05\textwidth]{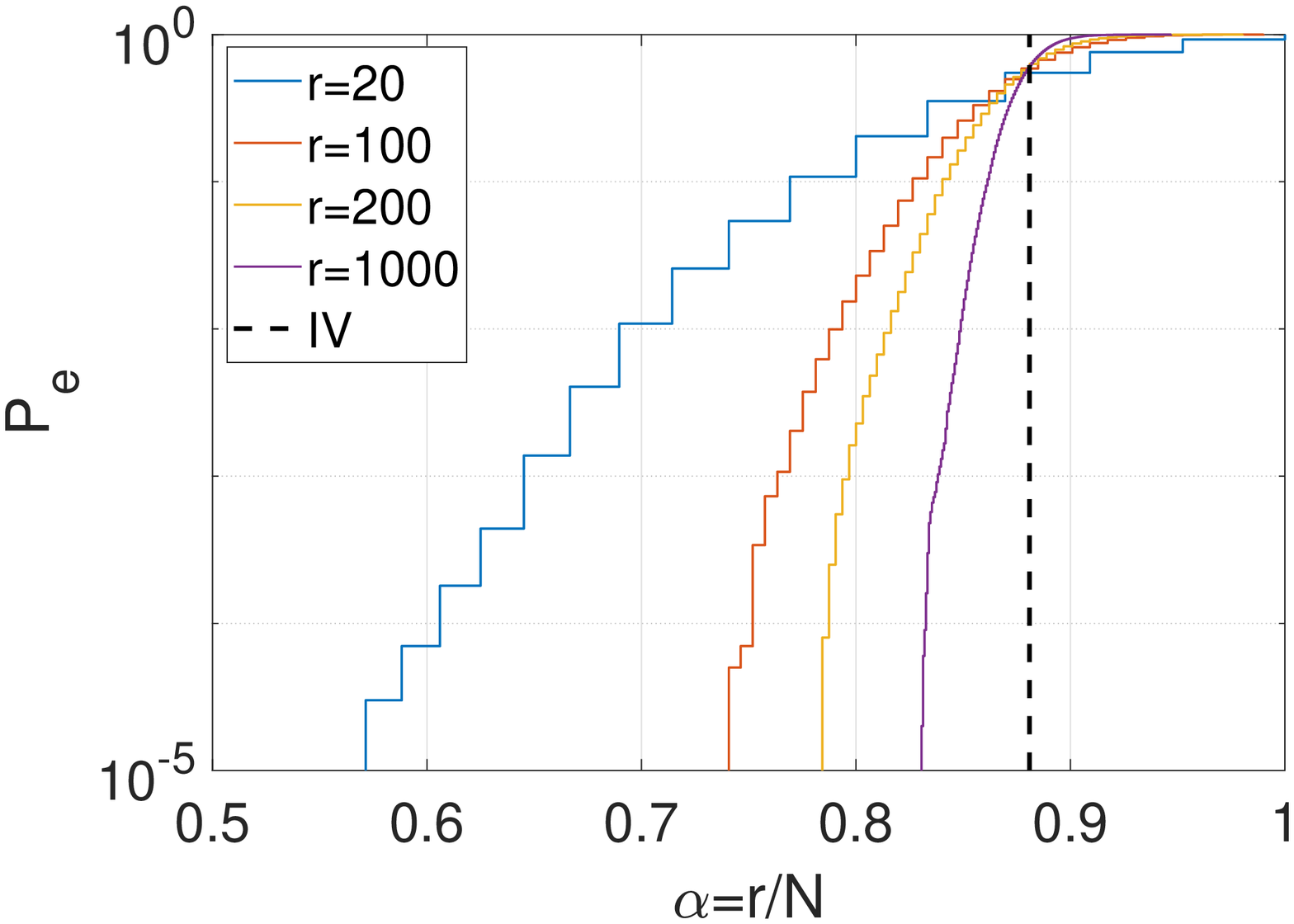}
        \caption{Gilbert--Eliott arrival process with $\gamma=0.01$, $\beta=0.1$, $\varepsilon=0.45$.}
    \label{fig:Hetero:GE-vs-r}
    \end{subfigure}
	\caption{Arrive-failure ratio as a function of $\alpha$ for $\lambda = 0.5$
	$\bP=(0.01, 0.1)$, $\bQ = (0.5, 0.5)$ for different number of relays $r$, and different arrival processes.}
    \label{fig:Hetero:vs-r}
\end{figure*}

Next, we examine the empirical arrive-failure ratio for \iid\ geometric interarrival times with rate $\lambda = 0.5$ and $p = 0.01$, for which $\IV = 0.98$.
\figref{fig:upper-lower-hom-diff-arrival-vs-N} depicts the empirical $P_e$ 
and the bounds on $P_e$ of \thmref{thm:IV:homogeneous} (with the extension of \corref{cor:lambda>0:all-results} for $\lambda > 0$)
against $N$ for $\alpha = 0.96$ and $p = 0.001$.
It shows that the lower bound is rather tight at least for these (and other) parameters.
\figref{fig:Hom-Arrival-Geo-vs-r} depicts the empirical $P_e$ and the lower bound on $P_e$ of \thmref{thm:IV:homogeneous} (with the extension of \corref{cor:lambda>0:all-results} for $\lambda > 0$) against $\alpha$ for several values of $r$; for each $r$, we change $\alpha = r/N$ by varying $N$. The simulations were carried for $10^6$ packets with only the last $10^5$ packets used for the evaluation, for which the system has reached steady state.
This figure nicely demonstrates that the arrive-failure probability approaches a step function at $\IV$ as the number of relays grows. Furthermore, the empiric curves concentrate around $P_e = 0.5$ at $\IV$.

Next, we depict in \figref{fig:Hetero:vs-r},
the empirical arrive-failure ratio for heterogeneous links in the fixed-type settings with 
$\bP = (0.01, 0.1)$ and $\bQ = (0.5, 0.5)$ for three different arrival processes, all of arrival rate $\lambda = 0.5$:
\begin{enumerate}[(a)]
\item
    Geometric \iid\ interarrival times with success (arrival) probability $0.5$.
\item 
    Deterministic interarrival times with an arrival every other time step.
\item 
    Gilbert--Eliott \cite{Gilbert:Gilbert-Eliott-model,Eliott:Gilbert-Eliott-model} arrival process with $\gamma = 0.01, \beta = 0.1, \veps = 0.45$---a two-state Markov model comprising a good state at which a packet arrives with probability $\veps$, and a bad state at which a new packet always arrives.
    $\gamma$ and $\beta$ are the transition probabilities from the good state to the bad state and vice versa.\footnote{The Gilbert--Eliott model was originally used to model bursts of errors. Here, we use this model for bursts of arrivals and hence the somewhat confusing state names.}
    The stationary probability of arrival equals $\frac{\beta}{\beta + \gamma} \cdot \varepsilon + \frac{\gamma}{\beta + \gamma}.$
\end{enumerate}
According to \thmref{thm:IV:hetero:fixed:delayed} with its extensions of Cors.~\ref{cor:lambda>0:all-results} and \ref{cor:lambda>0:all-results:non-geometric} to $\lambda > 0$, 
the IV is equal to $\IV \approx 0.881$. Again, the simulations were carried for $10^6$ packets with only the last $10^5$ packets used for the evaluation, for which the system has reached steady state.
All three arrival processes exhibit a similar behavior, with their respective empirical arrive-failure ratio curves approaching a step function at the $\IV$ as the number of relays increases. 
And again, as in the homogeneous case, the curves concentrate around $P_e = 0.5$ at $\IV$, for each of the arrival processes.


\section{Extensions and Related Settings}
\label{s:extensions}

\subsection{Alternative Definitions of the Information Velocity} 
\label{ss:extensions:IV:alternatives}

In this work, we have defined the IV as the maximum fraction of links that a packet may traverse with each additional time step such that the error probability decays to zero. This definition is important for setups of given number of relays and communication time. 
Furthermore, this definitions of the IV readily applies to 
physical channels (additive Gaussian noise channels, channels with errors, etc.); see also  \secref{ss:extensions:1packet:no-FB}.

Nevertheless, for packet-based communication other definitions of the IV are plausible, if one allows $r$ or $N$ to vary:
\begin{enumerate}
\item 
    Assume $N$ is given and one wishes to determine the \textit{expected} number of relays that a packet traverses within $N$ time steps. Then, the IV can be naturally defined as $\IV = \E{r}/N$. This definitions is appropriate for situation when one wishes to find the speed (or velocity) with which information spreads in a line network.
\item 
    Assume now the opposite case of a given number of relays $r$ that a packet needs to traverse, and the IV $\IV = r / \E{N}$. 
\end{enumerate}

Indeed, all three definitions, the two listed here and the original one of \eqref{eq:def:IV} coincide in the limit of $N \to \infty$. Moreover, the limit is not needed for the two alternative definitions for \iid\ geometric interarrival times (or for single-packet transmission).

\subsection{Heterogeneous Link-Type Patterns}

\begin{figure}[t]
	\vspace{-.1\baselineskip}
		\centering
	    \includegraphics[width=.85\columnwidth]{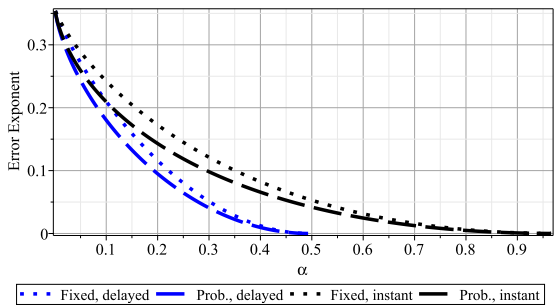}
	\caption{The EEs in natural base of Ths.~\ref{thm:IV:hetero:fixed:delayed} and \ref{thm:IV:hetero:prob:delayed} for $\bQ = \tbQ = (0.5, 0.2, 0.3)$ and $\bP = (0.2, 0.5, 0.7)$, along with their instantaneous counterparts.}
	\label{fig:EEs}
	\vspace{-.78\baselineskip}
\end{figure}

We have limited the number of possible erasure probabilities to $S < \infty$. 
However, the results of \thmref{thm:IV:hetero:prob:delayed} and their adaptation to $\lambda > 0$ of Cors.\ \ref{cor:lambda>0:all-results} and \ref{cor:lambda>0:all-results:non-geometric} may be extended to a continuum of possible erasure probabilities 
and probability density functions $\tbQ$, as well as for memory between the components of $\bp$, using large deviations theory \cite{Dembo-Zeitouni:LargeDeviations:Book}.
 Furthermore, the assumption that the sequence $\bp$ comprises \iid\ samples can be relaxed. For instance, the IV of \thmref{thm:IV:hetero:fixed:delayed} remains the same in settings where the portion of links having an erasure probability $P(i)$ converges to $Q(i)$: 
 $\lim_{r \to \infty} R(i)/r = Q(i)$;
 this is the case for 
 (positive) 
 recurrent Markov chains. For EE the situation is more subtle, since strong memory between the components of $\bp$ may alter the EE, as is implied by the difference between the EEs in the fixed-type and probabilistic settings in Ths.~\ref{thm:IV:hetero:fixed:delayed} and \ref{thm:IV:hetero:prob:delayed}, respectively; see \figref{fig:EEs}.

\subsection{Single-Packet Information Velocity Without Feedback} 
\label{ss:extensions:1packet:no-FB}

Polyanskiy (see \cite{Huleihel-Polyanskiy-Shayevitz:real-time-relaying:ISIT2019}; see also \cite{Rajagopalan-Schulman:FOCS:Real-Time-Networks:STOC1994}) defined the IV in a similar fashion to \eqref{eq:def:IV} for single-bit transmission over a cascade of binary symmetric channels without feedback. 

Interestingly, the proposed treatment in this work readily applies to single-packet (or bit) transmission over a cascade of binary erasure channels without feedback as well. To that end, define the error probability (\cf~\eqref{eq:failure-prob} for single-packet transmission)
\begin{align}
\label{eq:BLER:noFB}
    \BLER{N} \triangleq \PR{\hW(N) \neq W} ,
\end{align}
where the information packet $W$ is assumed to be uniform over $\{0, 1, \ldots, M-1\}$, 
$\hW(N)$ denotes the decoded packet at the end receiver after $N$ time steps.

Then, the following policy at each node is optimal.\footnote{\label{foot:non-unique} Clearly, the optimal policy is not unique.}
\begin{scheme*} 
\ 

    \textit{Transmitter:} 
    Repeats the source packet $W$ at all times.

    \textit{Node $i \in [2:r]$ at time $t \in \nats$:} 
    \begin{itemize} 
    \item 
        If it observed a non-zero packet at some time $\ell \in [t]$,
        sends that packet.
    \item 
        Otherwise (if it observed only erasures and zero measurements), sends $0$.
    \end{itemize} 

    \textit{End receiver at time $t$:} If it observed a non-zero packet at some time $\ell \in [t]$, declares $\hW(t)$ as that packet; otherwise, decodes $\hW[t] = 0$.
\end{scheme*}
Clearly, $P_e$ of \eqref{eq:BLER:noFB} is the same as that of \eqref{eq:failure-prob} of the single-packet transmission setting of \secref{s:1packet} up to a multiplicative factor of $1 - M^{-1}$. Thus, all the results of \secref{s:1packet} hold true for the scenario described in this subsection up to this multiplicative factor in the error/arrive-failure probability expressions and bounds, which has no bearing to the IV and EE results. 

\subsection{Instantaneous Links} 
\label{ss:extensions:1packet:instant-hops}

In this work, we assumed that each link causes a delay of at least one time unit. 
Yet, one may consider a setup in which the links are ``instantaneous'', in which no such delay is incurred. 
Although this seems like an ill-modeled setup in the presence of ACKs, it might be plausible under the framework of \secref{ss:extensions:1packet:no-FB}.

For this setup, $\tau_i$ in \eqref{eq:delayed:BLER:taus} are distributed according to a non-shifted geometric distribution, \ie, a distribution that counts only the failures until the first success \textit{not including} the success (\cf.~\remref{rem:geometric-pdf}).
Thus, the bounds in \eqref{eq:thm:IV:homo:failure-prob} of \thmref{thm:IV:homogeneous} hold true by replacing $N$ with $N + r$ [which for a fixed $\alpha$, is equal to $(1+\alpha) N$ up to rounding errors], and the adaptation of all the results of \secref{s:1packet} is straightforward. In particular, the IV and the EE in the homogeneous-links setting become $\IV = \frac{1-p}{p}$ and $\Eblk = (1 + \alpha) \KL{\frac{\alpha}{1 + \alpha}}{1 - p}$, respectively. Note that $\IV > 1$ for $p < 1/2$, meaning that more than $N$ (instantaneous) links may be traversed within $N$ time steps. 

We compare the EE of \thmref{thm:IV:hetero:fixed:delayed} with
channels-type $\bQ = (0.5, 0.2, 0.3)$, and the EE of \thmref{thm:IV:hetero:prob:delayed} for $\tbQ = (0.5, 0.2, 0.3)$, both for $\bP = (0.2, 0.5, 0.7)$; for these parameters, $\IV \approx 0.49$.
We further compare them to their instantaneous counterparts, for which $\IV \approx 0.98$. The four EEs are depicted in \figref{fig:EEs}.


\section{Discussion} 
\label{s:discussion}

\subsection{Single Packet Over $r$ Links vs.\ $r$ Packets Over One Link}
\label{ss:discussion:1packet-r-links_vs_r-packets-1link}

The setting of \secref{ss:IV:same-channel} of single-packet transmission over a cascade of $r$ homogeneous links across $N$ time steps is mathematically equivalent to transmitting $r$ packets over a single link (with feedback) during $N$ time steps.\footnote{More generally, one may show that transmitting $m$ packets over $r$ links is mathematically equivalent to transmitting $r$ packets over $m$ links during the same amount of time.}
Since the capacity of the latter is known to be $1 - p$ packets per time unit (with or without feedback, since the link is memoryless), this view allows to recover the IV of \thmref{thm:IV:homogeneous}. 

However, for the two settings for heterogeneous links of \secref{ss:IV:different-channels}, the analogy becomes somewhat peculiar:
Consider the probabilistic setup of \thmref{thm:IV:hetero:prob:delayed}. In the analogous $r$-packets single-link setting, 
the erasure probability is chosen independently at the beginning of transmission (service) of each packet, but remains fixed during the entire transmission of this packet and until it is successfully conveyed to the receiver. 
The capacity of this setting (which is equal to the IV of \thmref{thm:IV:hetero:prob:delayed}) is different from the capacity of a link with \iid\ erasures across all time steps (in-fact all packets), regardless of the success or failure events in previous time steps, which is equal to $1 - \sum_i Q(i) P(i) = \sum_i (1 - Q(i)) P(i)$.

In fact, by simple convexity arguments the capacity of the $r$-packets--single-link setting can be shown to be lower than the that of single-packet--$r$-links setting. 
The explanation for this is reminiscent of the waiting-time paradox: In the former case, once a large erasure probability is selected it takes a long time until a success, i.e., until a low erasure probability is selected, and vice versa; whereas in the latter case, no such extra loss is incurred.

For a stream of packets, however, this analogy breaks down, since the problem becomes two-dimensional by its nature with respect to the arriving packets and to the traversed links.

\subsection{Continuous-Time Processes}

Since continuous-time processes can be viewed as limit processes of discrete-time ones, the derived results in this work can be adapted to
a continuous-time setting with exponential transmission time over each link. In fact, all the results from queueing theory that were used in \secref{s:lambda>0} were originally derived for continuous-time processes.

\subsection{In-Order Transmission Assumption}

We have assumed in-order transmission in all the relays. This setting is appropriate when the relays do not know the target delay or the number of relays to the end destination. However, for better informed relays, breaking the in-order assumption may improve performance. 
To illustrate this, consider the homogeneous-links setting with $ 1 - p < \lambda < 2(1 - p)$, \ie, a setting in which the stable-queues assumption of \eqref{eq:lambda<mu} is violated, with the second assumption being technical but useful for this illustration. In this situation the number of packets in each of the buffers grows indefinitely, meaning that the resulting arrive-failure probability goes to 1. However, by dropping every other packet at the transmitter (i.e, transmitting only every other packet), one reduces the effective arrival rate of the remaining packets to $\lambda/2$; by our technical assumption these transmitted packets satisfy the stability condition \eqref{eq:lambda<mu} and by \thmref{thm:IV:homogeneous} have a vanishingly small error of arrive-failure of probability for large values of $N$. Thus, the overall average arrive-failure probability is reduced to approximately $1/2$ instead of $1$. Of course, more sophisticated policies may be utilized by the different nodes, which compare, for example, the accumulated delay of the different packets in each buffer, and giving lower priority to packets that are likely to fail to arrive to the destination within their allowed delay. We leave this direction for future research.

\subsection{Anytime Anywhere Reliability}

The exponential decay of the arrive-failure probability with the elapsed time (delay) of each packet individually may be reinterpreted as 
the \textit{anytime reliability} property of Sahai and Mitter \cite{SahaiMitterPartI,Sahai:NotTheSame:IT2008} (see also \cite{CausalPM:ISIT2019,SukhavasiHassibi,TreeCodes:ISIT2016}) that is fundamental in networked control systems, 
with the EE taking the role of \textit{anytime EE} of Sahai and Mitter, \ie, determining the plant unstable eigenvalues that can be stabilized. Moreover, in our setting of a cascade of multiple links, one may further discuss anytime \textit{anywhere} reliability by viewing the arrive-failure probability at different start and end nodes.
It would be also interesting to extend these results to the setting without feedback \cite{SahaiMitterPartI} (see also \cite{SukhavasiHassibi,TreeCodes:ISIT2016,TreeCodes}); this is currently under intensive scrutiny. 

\subsection{General Channels and Networks}

It would be interesting to extend the IV notion to more general communication links 
\cite{Ling-Scarlett:Real-Time-Relaying:ISIT2021,Ling-Scarlett:Real-Time-Relaying:Arxiv2021} 
and more general networks \cite{Rajagopalan-Schulman:FOCS:Real-Time-Networks:STOC1994,Information-Velocity:Wireless:WiOpt2015}. 
A recent effort in this direction was made in \cite{Ling_Scarlett:IV:2021}.


\section*{Acknowledgment}

We thank Yury Polyanskiy for introducing us to the problem of information velocity, and Anant Sahai for an interesting discussion about anytime reliability with feedback. 

\newpage
\bibliographystyle{IEEEtran}
\bibliography{toly1}

\end{document}